\documentclass[preprint,authoryear,12pt]{elsarticle}
\usepackage{tikz}
\usetikzlibrary{arrows, arrows.meta, positioning}
\usepackage{fullpage}
\usepackage{graphicx}
\usepackage{xcolor}
\usepackage{float}
\usepackage{hyperref}
\usepackage{subcaption}
\usepackage[export]{adjustbox}
\usepackage[bottom]{footmisc}
\usepackage{bm}
\usepackage[margin=2.5cm]{geometry}
\usepackage{amsmath,amsthm,amssymb}
\usepackage{textcomp}
\usepackage{graphicx,psfrag}
\newcommand{\be}{\begin{equation}}
\newcommand{\en}{\end{equation}}

\def\bm#1{\mbox{\boldmath{$#1$}}}

\numberwithin{equation}{section}

\definecolor{clr1}{RGB}{235,240,252}
\definecolor{pr}{RGB}{255,105,97}
\newcommand\mc{\mathcal}

\newcommand\tx{\text}
\newcommand\Ri{R_\text{i}}
\newcommand\Ro{R_\text{o}}

\newcommand\gcr{\gamma_{\tx{cr}}}
\newcommand\Jm{J_{\tx{m}}}

\newcommand\RA{\rightarrow}
\newcommand\spc{\,\,\,\,\,\,\,\,}
\newcommand\pd{\partial}

\newcommand{\lz}{\lambda_{z}}
\newcommand{\lt}{\lambda_{\theta}}
\newcommand{\lzcr}{\lambda_{z\tx{cr}}}
\usepackage[font=footnotesize]{caption}
\usepackage{titlesec}

\long\def\symbolfootnote[#1]#2{\begingroup\def\thefootnote{\fnsymbol{footnote}}\footnote[#1]{#2}\endgroup}

\def\bm#1{\mbox{\boldmath{$#1$}}}

\begin{document}
\begin{frontmatter}

\title{\textsf{\textbf{Elasto-capillary necking, bulging and Maxwell states in soft compressible cylinders.}}}

\author[mymainaddress]{Dominic Emery\corref{mycorrespondingauthor}}
\cortext[mycorrespondingauthor]{Corresponding author}
\ead{d.r.emery@keele.ac.uk}
\address[mymainaddress]{School of Computing and Mathematics, Keele University, Staffordshire ST5 5BG, UK}

\begin{abstract}
\noindent Localized pattern formations and ``two-phase" deformations are studied theoretically in soft compressible cylinders subject to surface tension and axial loading through several force-controlled loading scenarios. By drawing upon known results for separate yet mathematically similar elastic localization problems, a concise family of analytical bifurcation conditions for localized bulging or necking are derived in terms of a general compressible strain energy function. The effect of material compressibility and strain-stiffening behaviour on the bifurcation point is analysed, and comparisons between our theoretical bifurcation conditions and the corresponding numerical simulation results of Dortdivanlioglu and Javili (Extreme Mech. Lett. \textbf{55}, 2022) are made. It is then explained how the fully developed ``two-phase" (Maxwell) state which evolves from the initial localized bifurcation solution can be comprehensively understood
 using the simple analytical expressions for the force parameters corresponding to the primary axial tension deformation. The power of this simple analytical approach in validating numerical simulation results for elastic localization and phase-separation-like problems of this nature is highlighted.
\end{abstract}


\end{frontmatter}


\section{Introduction}

The treatment of localized pattern formation in solid cylinders and hollow tubes as a bifurcation problem has become increasingly prevalent over the last 15 years. A problem which serves as the foundation for this area of research is the localized bulging of a hollow tube subject to the combined effects of axial loading and internal inflation. Despite many experimental observations in the past \citep{mallock1891,kyriakides1990}, this was only recognized as a bifurcation phenomenon with zero wavenumber under the membrane assumption relatively recently by \cite{fu2008}. \cite{fu2016} demonstrated that, for a tube of \textit{arbitrary thickness}, the bifurcation condition for localized bulging is that the Jacobian determinant of the inflation pressure $P$ and the resultant axial force $\mc{N}$ as functions of the axial stretch and the circumferential stretch on the inner surface must vanish. This is equivalently the condition for an axi-symmetric bifurcation mode with zero wavenumber to exist \citep{yu2022}. Previous linear bifurcation analyses \citep{ho1979a,ho1979b} had focussed on periodic axi-symmetric modes, and the zero wavenumber mode was incorrectly thought to correspond to an alternate uniformally inflated state. Since the revelation of \cite{fu2016}, many additional effects such as rotation \citep{wang2017}, double fibre-reinforcement \citep{wang2018effect}, bi-layering \citep{liu2019} and torsion \citep{althobaiti2022} have been incorporated into the analysis. 

The inflation problem has become prototypical in the sense that it often has a very similar mathematical structure to other more complicated elastic localization problems. For instance, through a reformulation of the Jacobian determinant bifurcation condition for the inflation problem, \cite{fu2018} demonstrated that the bifurcation condition for localized necking in a dielectric membrane under in-plane mechanical stretching and an electric field is that the Hessian of the total free-energy function vanishes. The problem of localized bulging or necking in soft incompressible cylinders and tubes under axial loading and surface tension has also become very well understood as a result of the inflation problem.

Surface tension operates in solids at the elasto-capillary length scale $\gamma/\mu$, where $\gamma$ is the surface tension and $\mu$ is the ground-state shear modulus \citep{bico2018}. For extremely soft materials such as gels, creams and biological tissue, surface tension effects can be of the same order of magnitude as the bulk elastic modulus at length scales ranging from tens of nanometres to millimetres. Thus, when modelling the large deformations of these small-scale materials, surface tension effects cannot be neglected. The localized axi-symmetric ``beading" of soft slender cylinders, such as axons, has been widely observed experimentally \citep{matsuo1992,bar1994,fong1999}. The implication of this phenomenon in nerve damage due to traumatic brain injuries \citep{kilinc2009} and neurodegenerative conditions such as Alzheimer's and Parkinson's diseases \citep{datar2019} has motivated many recent studies surrounding the bifurcation behaviour of an incompressible solid cylinder under a resultant axial force $\mc{N}$ and a surface tension $\gamma$.

The aforementioned elasto-capillary problem was initially studied using non-linear elasticity theory by \cite{taffetani2015,taffetani2015a} and \cite{xuan2016}. However, through a weakly non-linear analysis, \cite{FuST} was the first to demonstrate that the initial bifurcation is a sub-critical localized necking or bulging solution (depending on the nature of the loading), and this is again associated with zero axial wavenumber. The corresponding bifurcation condition was also shown to take a simple analytical form in this case; for fixed $\mc{N}$ (fixed $\gamma$) and increasing $\gamma$ (varying $\mc{N}$), localized necking or bulging occurs when $d\gamma/d\lz =0$ ($d\mc{N}/d\lz = 0$), where $\lz$ is the axial stretch. \cite{xuan2017} and \cite{giudici2020} highlighted that the complete bifurcation process is a \textit{phase-transition-like} phenomenon which culminates in a ``two-phase" state consisting of two regions of distinct but uniform axial stretch $\lz$ connected by a smooth transition zone. The connection between the initial localized bifurcation solution and the final ``two-phase" state was explained both theoretically and through numerical simulations by \cite{FuST}. The beading instability in solid cylinders has since been studied dynamically \citep{pandey2021} and through
the active strain approach \citep{riccobelli2021}. The observations of \cite{FuST} for a solid cylinder have also been fully extended to the case of an incompressible hollow tube by \cite{emery2021IJSS,emery2021PRSA}. Circumferential buckling instabilities in cylinders and tubes under surface tension and axial loading \citep{emery2021MOSM}, growth \citep{bevilacqua2020}, and uniform pressure and geometric everting \citep{wang2021large} have also been extensively studied in recent years.

The case of a \textit{compressible} solid cylinder has received very little attention in the literature. This is surprising since, whilst the incompressibility assumption typically makes the bifurcation analysis far easier, soft hydrogels can often possess a large degree of compressibility. Furthermore, in the case of soft biological tissue, whilst incompressibility is often assumed due to the high water content of the material, there is very little supporting experimental evidence for this assumption. Only \cite{carew1968} has provided evidence that incompressibility is a suitable assumption in modelling arterial tissue. Very recently, however, \cite{dort2022} (hereafter abbreviated as ``DJ") analysed the effect of material compressibility via numerical simulations, extending what is already known for the incompressible case. Numerical simulation predictions for the initial bifurcation points are presented for two separate compressible neo-Hookean strain energy functions, and an extensive post-bifurcation analysis tracking the axial propagation of the localized solution is performed. The work offers a different numerical perspective on the existing literature for the incompressible case, but does not present any analytical results to compare with the numerical predictions for the compressible case. The aim of this paper is two-fold. Firstly, we look to extend the known analytical bifurcation conditions of \cite{FuST} to the compressible case for a general strain energy function, and make comparisons with the numerical bifurcation conditions in DJ. Secondly, we wish to highlight how the \textit{entire} post-bifurcation process can be captured through analytical means, and it is hoped that this will encourage future studies to use analytical approaches to guide and corroborate numerical simulation results.

The remainder of this paper is organised as follows. In the next section, we present the primary axial tension deformation and derive corresponding analytical expressions for the resultant axial force $\mc{N}$ and the surface tension $\gamma$. In section $3$, analytical bifurcation conditions for localized bulging or necking are derived in terms of a general strain energy function for several force-controlled loading scenarios. The effect of compressibility and strain-stiffening behaviour on the bifurcation point is considered, and comparisons are made with the numerical results of DJ. In section $4$, we highlight a powerful analytical approach to describing the full post-bifurcation process of the cylinder, and make comparisons with the corresponding results in DJ. Concluding remarks are finally given in section $5$.


\section{Primary deformation}
Consider a \textit{compressible}, isotropic, hyperelastic solid cylinder with a reference configuration $\mc{B}_{0}$ defined in terms of the cylindrical polar coordinates $(R,\Theta,Z)$, where
\begin{align}
0\leq R\leq\Ro,\spc 0\leq\Theta\leq 2\pi,\spc |Z|<L.
\end{align}
The finitely deformed configuration $\mc{B}_{e}$ is in terms of the cylindrical polar coordinates $(r,\theta,z)$, and we assume that the solid cylinder undergoes a primary homogeneous deformation of the form
\begin{align}
r=\lambda_{\theta}R,\spc\theta=\lt\Theta,\spc z=\lz Z,
\end{align}
where $\lt$ and $\lz$ are the constant circumferential and axial stretches, respectively. Thus, we have that
\begin{align}
0\leq r\leq\lt\Ro,\spc 0\leq\theta\leq 2\pi,\spc |z|<\lz L.
\end{align}
The position vectors of a representative material particle in $\mc{B}_{0}$ and $\mc{B}_{e}$ are given respectively by
\begin{align}
\bm{X}=R\bm{E}_{R}+Z\bm{E}_{Z},\spc\bm{x}=r\bm{e}_{r}+z\bm{e}_{z},
\end{align}
where $(\bm{E}_{R},\bm{E}_{\Theta},\bm{E}_{Z})$ and $(\bm{e}_{r},\bm{e}_{\theta},\bm{e}_{z})$ are the corresponding orthonormal bases. The primary deformation gradient $F$ is then defined through $d\bm{x}=F d\bm{X}$ and may be written as
\begin{align}
F=\lt\left(\bm{e}_{r}\otimes\bm{E}_{R} + \bm{e}_{\theta}\otimes\bm{E}_{\Theta}\right) + \lz\bm{e}_{z}\otimes\bm{E}_{Z}.
\end{align}
Given the one-to-one correspondence between $\mc{B}_{0}$ and $\mc{B}_{e}$, we must have that $J\equiv\det F>0$. The associated left Cauchy-Green strain tensor $B=FF^{T}$ takes the form
\begin{align}
B=\lt^{2}\left(\bm{e}_{r}\otimes\bm{e}_{r} + \bm{e}_{\theta}\otimes\bm{e}_{\theta}\right) + \lz^{2}\,\bm{e}_{z}\otimes\bm{e}_{z}, \label{Bc6}
\end{align}
and its three principal invariants are expressed through
\begin{align}
I_{1}=\tx{tr}B=2\lt^{2}+\lz^{2},\spc I_{2}=\frac{1}{2}\left(I_{1}^{2}-\tx{tr}B^2\right)=\lt^{2}\left(2\lz^{2}+\lt^{2}\right),\spc I_{3}=J^2=\lambda_{\theta}^{4}\lz^{2}. \label{invsss}
\end{align}

We assume that the constitutive behaviour of the material is governed by a strain energy function of the form
\begin{align}
W=W(I_{1},I_{3}). \label{wIBJ}
\end{align}
In the computation of our results, we will predominantly specify $W$ to the following compressible Gent material model in order to account for strain-stiffening behaviour:
\begin{align}
W=-\frac{\mu}{2}\left\{\Jm\ln\left(1-\frac{I_{1}-3}{\Jm}\right)+2\ln J\right\} + \frac{1}{2}\lambda\left\{\frac{1}{2}(J^2 -1) - \ln J\right\}. \label{gentC}
\end{align}
The parameter $\Jm$ is the material extensibility limit, $\mu$ is the ground-state shear modulus and $\lambda=2\nu/(1-2\nu)$, where $\nu\in[0,1/2]$ is Poisson's ratio. For a fully compressible material, we have that $\nu\RA 0$, whilst the incompressible limit corresponds to $\nu\RA 1/2$. In order to facilitate a comparison between our theory and the numerical results of DJ, we will also consider the quadratic and logarithmic compressible neo-Hookean material models given respectively by:
\begin{flalign}
&& &W=\frac{1}{2}\mu\,(I_{1}-3-2\log J) + \frac{1}{2}\lambda\left\{\frac{1}{2}(J^2 -1) - \log J\right\},\label{WQ}& \\[0.4em] \text{and}
&& &\spc\spc\,\,\,\, W=\frac{1}{2}\mu\,(I_{1}-3-2\log J) + \frac{1}{2}\lambda\,(\log J)^{2}\label{WL}.&
\end{flalign}
Note that, in the limit $\Jm\RA\infty$, the compressible Gent material model $(\ref{gentC})$ reduces to the quadratic neo-Hookean model $(\ref{WQ})$. All of the algebraic manipulations and computations associated with the following work have been performed in \textit{Mathematica} \citep{wo2021}.


\subsection{Stress-based formulation}
Under the assumption $(\ref{wIBJ})$, the Cauchy stress tensor $\sigma$ is expressible as
\begin{align}
\sigma = 2JW_{3} I + 2J^{-1}W_{1}B, \label{sig}
\end{align}
where the notation $W_{i}=\pd W/\pd I_{i}$ and $W_{ij}=\pd^2 W/\pd I_{i} \pd I_{j}$ for $i,j=1,2,3$ is employed here and hereafter, and $I$ is the identity tensor. On substituting $(\ref{Bc6})$ into $(\ref{sig})$, the Cauchy stresses in the radial, circumferential and axial directions are found to take the form
\begin{gather}
\sigma_{rr}=\sigma_{\theta\theta}=2\lt^2 \lz W_{3} + 2\lz^{-2}W_{1},\spc\sigma_{zz}=2\lt^2 \lz W_{3} + 2\lt^{-2}W_{1}. \label{sigcomps}
\end{gather}
These components are constant, and so the equilibrium equations $\tx{div}\,\sigma=\bm{0}$ are automatically satisfied. The solid cylinder is under the combined effect of a surface tension $\gamma$ and a resultant axial force $\mc{N}$. We scale all lengths by $\Ro$, all stresses by the ground state shear modulus $\mu$ and the surface tension $\gamma$ by $\mu\Ro$. As such, $\Ro$ and $\mu$ may be set equal to unity without loss of generality; we use the same symbols to denote scaled quantities.

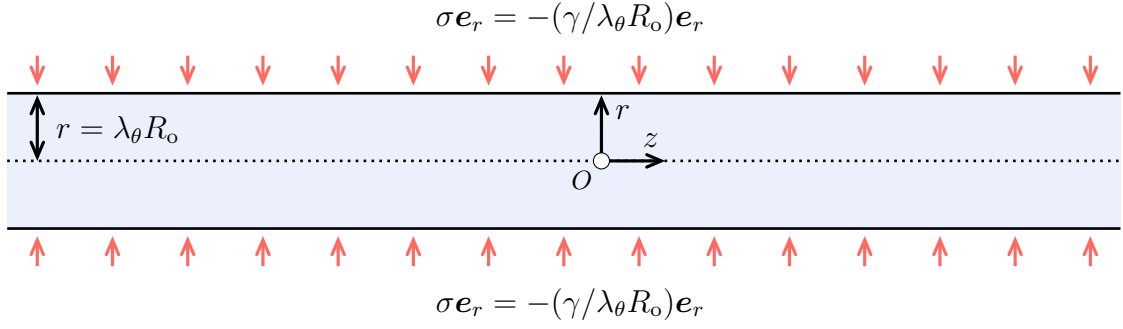
\begin{figure}[h!]
\centering
\begin{tikzpicture}
\path [fill=clr1] (0.1,0.1) rectangle (14.9,1.9);
\draw [line width=0.35mm] (0.1,1.9)--(14.9,1.9);
\draw [line width=0.35mm] (0.1,0.1)--(14.9,0.1);
\draw [line width=0.35mm,dotted] (0.1,1)--(14.9,1);
\draw[>=angle 45, pr, ->, very thick] (0.5,2.4) -- (0.5,2);
\draw[>=angle 45, pr, ->, very thick] (14.5,2.4) -- (14.5,2);
\draw[>=angle 45, pr, ->, very thick] (1.5,2.4) -- (1.5,2);
\draw[>=angle 45, pr, ->, very thick] (13.5,2.4) -- (13.5,2);
\draw[>=angle 45, pr, ->, very thick] (2.5,2.4) -- (2.5,2);
\draw[>=angle 45, pr, ->, very thick] (12.5,2.4) -- (12.5,2);
\draw[>=angle 45, pr, ->, very thick] (3.5,2.4) -- (3.5,2);
\draw[>=angle 45, pr, ->, very thick] (11.5,2.4) -- (11.5,2);
\draw[>=angle 45, pr, ->, very thick] (4.5,2.4) -- (4.5,2);
\draw[>=angle 45, pr, ->, very thick] (10.5,2.4) -- (10.5,2);
\draw[>=angle 45, pr, ->, very thick] (5.5,2.4) -- (5.5,2);
\draw[>=angle 45, pr, ->, very thick] (9.5,2.4) -- (9.5,2);
\draw[>=angle 45, pr, ->, very thick] (6.5,2.4) -- (6.5,2);
\draw[>=angle 45, pr, ->, very thick] (8.5,2.4) -- (8.5,2);
\draw[>=angle 45, pr, ->, very thick] (7.5,2.4) -- (7.5,2);
\draw[>=angle 45, pr, ->, very thick] (0.5,-0.4) -- (0.5,0);
\draw[>=angle 45, pr, ->, very thick] (14.5,-0.4) -- (14.5,0);
\draw[>=angle 45, pr, ->, very thick] (1.5,-0.4) -- (1.5,0);
\draw[>=angle 45, pr, ->, very thick] (13.5,-0.4) -- (13.5,0);
\draw[>=angle 45, pr, ->, very thick] (2.5,-0.4) -- (2.5,0);
\draw[>=angle 45, pr, ->, very thick] (12.5,-0.4) -- (12.5,0);
\draw[>=angle 45, pr, ->, very thick] (3.5,-0.4) -- (3.5,0);
\draw[>=angle 45, pr, ->, very thick] (11.5,-0.4) -- (11.5,0);
\draw[>=angle 45, pr, ->, very thick] (4.5,-0.4) -- (4.5,0);
\draw[>=angle 45, pr, ->, very thick] (10.5,-0.4) -- (10.5,0);
\draw[>=angle 45, pr, ->, very thick] (5.5,-0.4) -- (5.5,0);
\draw[>=angle 45, pr, ->, very thick] (9.5,-0.4) -- (9.5,0);
\draw[>=angle 45, pr, ->, very thick] (6.5,-0.4) -- (6.5,0);
\draw[>=angle 45, pr, ->, very thick] (8.5,-0.4) -- (8.5,0);
\draw[>=angle 45, pr, ->, very thick] (7.5,-0.4) -- (7.5,0);
\draw[>=angle 45, <-, very thick] (8,1.875) -- (8,1);
\draw[>=angle 45, <-, very thick] (8.85,1) -- (8,1);
\draw[>=angle 45, <->, very thick] (0.5,1) -- (0.5,1.875);
\node at (8,1) [circle,fill=white,draw=black,inner sep=2.15pt]{};
\node[text width=4cm] at (7.8,2.9)
    {$\sigma\bm{e}_{r}=-(\gamma/\lambda_{\theta}\Ro)\bm{e}_{r}$};
\node[text width=4cm] at (7.8,-0.9)
    {$\sigma\bm{e}_{r}=-(\gamma/\lambda_{\theta}\Ro)\bm{e}_{r}$};
\node[text width=1cm] at (8.685,1.65)
    {$r$};
\node[text width=1cm] at (9.05,1.2775)
    {$z$};
\node[text width=0.5cm] at (7.85,0.75)
    {\footnotesize $O$};
\node[text width=2cm] at (1.75,1.415)
    {$r=\lt\Ro$};
\end{tikzpicture}
\vspace{5mm}
\caption{A schematic of the current configuration $\mc{B}_{e}$ of the solid cylinder in the $(z,r)$ plane and the associated boundary conditions on the lateral surface $r=\lambda_{\theta}\Ro$.}
\label{SCschem}
\end{figure}

The surface tension enters the analysis through the boundary condition
\begin{align}
\sigma_{rr}=-\gamma/(\lambda_{\theta}\Ro),\spc r=\lambda_{\theta}\Ro; \label{bc}
\end{align}
see Fig. $\ref{SCschem}$.
On substituting $(\ref{sigcomps})_{1}$ into $(\ref{bc})$, we obtain the following expression for $\gamma$ in terms of $\lambda_{\theta}$ and $\lz$:
\begin{align}
\gamma=-2\lt\Ro\left\{\lt^{2}\lz W_{3} + 2\lz^{-2}W_{1}\right\}. \label{compgamrel}
\end{align}
The resultant axial force $\mc{N}$ is defined through
\begin{align}
\mc{N}&=\int_{\theta=0}^{2\pi}\int_{r=0}^{\lt\Ro}\sigma_{zz}rdrd\theta + \gamma\int_{\theta=0}^{2\pi}\lt\Ro d\theta\nonumber\\[0.4em]
&=2\pi\lt\Ro\left\{\Ro\lt^{3}\lz W_{3} + \Ro\lt^{-1}W_{1} + \gamma\right\}. \label{compN}
\end{align}


\section{Bifurcation conditions for localization}
We are interested in the bifurcation behaviour of the finitely deformed configuration $\mc{B}_{e}$. More specifically, we wish to determine the critical value of the load parameter at which localized patterns such as necking or bulging become theoretically possible. We consider three distinct types of loading which may be summarised as follows.

\begin{enumerate}
\item[\textit{(1)}] \textit{Fixed $\gamma$ and varying $\mc{N}$:}\\
A fixed surface tension $\gamma>0$ is applied to the cylinder which is also initially under zero resultant axial force $\mc{N}$ or a large strictly positive resultant axial force (due to the application of a dead weight at an end of the tube, say). In the former case, fixed surface tension produces an axial compression such that $\lz <1$ initially, and from this point we may increase $\mc{N}$ monotonically from zero to trigger bifurcation. In the latter case, we can instead decrease $\mc{N}$ monotonically from its large starting point determined by the dead weight in order to incite a bifurcation. We refer to these cases as axial force controlled ``loading" and ``unloading", respectively.

\item[\textit{(2)}] \textit{Fixed $\lz$ and increasing $\gamma$:}\\
The total length of the tube is fixed both before and after bifurcation into a localized necking or bulging solution has taken place. In other words, the averaged axial stretch $\lz\geq 1$, defined generally as the deformed length of the cylinder divided by the undeformed length, is fixed. The surface tension $\gamma$ is then increased monotonically from zero to prompt a bifurcation.

\item[\textit{(3)}] \textit{Fixed $\mc{N}$ and increasing $\gamma$:}\\
The resultant force $\mc{N}\geq 0$ is fixed, and this induces an initial axial stretch $\lz\geq 1$. The surface tension $\gamma$ is then increased monotonically from zero to trigger bifurcation, and the axial stretch $\lz$ will decrease monotonically from its aforementioned starting value in tandem to preserve the constant value of $\mc{N}$.
\end{enumerate}
\subsection{Fixed $\gamma$ and varying $\mc{N}$}
For any fixed $\gamma\geq 0$, we can define the circumferential stretch $\lt$ as an implicit function of the axial stretch $\lz$ through $(\ref{compgamrel})$. Based on the analysis of \cite{FuST} and \cite{emery2021IJSS,emery2021PRSA} for the incompressible solid cylinder and hollow tube case (respectively), we may then conjecture that the bifurcation condition for elasto-capillary localization is $d\mc{N}/d\lz = 0$, where $\mc{N}$ is given in $(\ref{compN})$ and $\gamma$ is fixed in the differentiation. This bifurcation condition takes the form
\begin{align}
\gamma&=-\frac{2\lz}{\lambda_{\theta\tx{d}}}W_{11}-\frac{\lt^{4}}{\lambda_{\theta\tx{d}}}\left\{W_{3} + 2\lz(\lz +1)W_{13} + 2\lt^{4}\lz^{2}W_{33}\right\}  \nonumber\\[0.4em]
&\spc -4\lt\left\{ W_{11} + \lt^{2}\lz W_{3} +\lt^{2}\lz(\lz + \lt^{2})W_{13} + \lt^{6}\lz^{3}W_{33}\right\}, \label{bccompr}
\end{align}
where $\lambda_{\theta\tx{d}}=d\lambda_{\theta}/d\lz$, and the right-hand side of $(\ref{bccompr})$ is evaluated at the critical value of $\lz$, $\lzcr$. By differentiating equation $(\ref{compgamrel})$ implicitly with respect to $\lz$, the following explicit expression for $\lambda_{\theta\tx{d}}$ in terms of $\lz$ and $\lt$ can be obtained:
\begin{align}
\lambda_{\theta\tx{d}} =\lt\frac{4W_{1} - \lz^{2}\left\{4 W_{11} +\lt^{2}\lz W_{3}+ 2\lambda_{\theta}^6 \lz^{3} W_{33} + 2\lambda_{\theta}^{2}(\lz^3 + 2\lambda_{\theta}^2)W_{13}\right\}}{2\lz W_{1} + 8\lambda_{\theta}^{2}\lz W_{11} + \lambda_{\theta}^2 \lz^3 \left\{3\lz W_{3} + \lambda_{\theta}^{4}\lz^3 W_{33} + 4\lambda_{\theta}^{2}(\lz +2)W_{13}\right\}}. \label{ltdd}
\end{align}
For a given fixed $\gamma$, the bifurcation values $\lzcr$ can be determined numerically from $(\ref{bccompr})$. Then, with use of $(\ref{compN})$, the associated bifurcation values of the resultant axial force, $\mc{N}_{\tx{cr}}\equiv\mc{N}(\lzcr)$, may be computed.

\begin{figure}[h!]
\centering
\begin{tikzpicture}
\node at (0,0) {\includegraphics[scale=0.38]{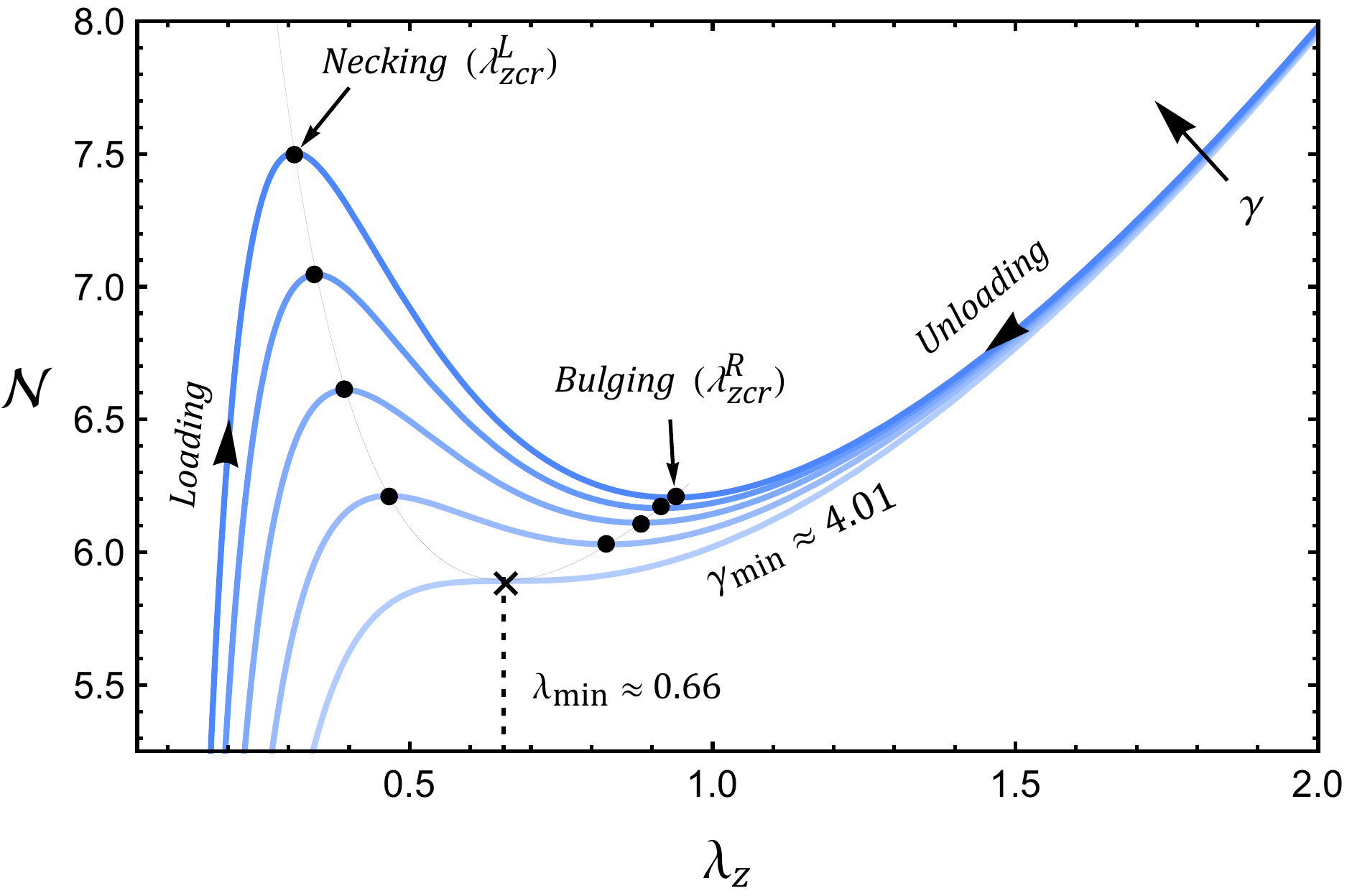}};
\node at (7.5,0) {\includegraphics[scale=0.375]{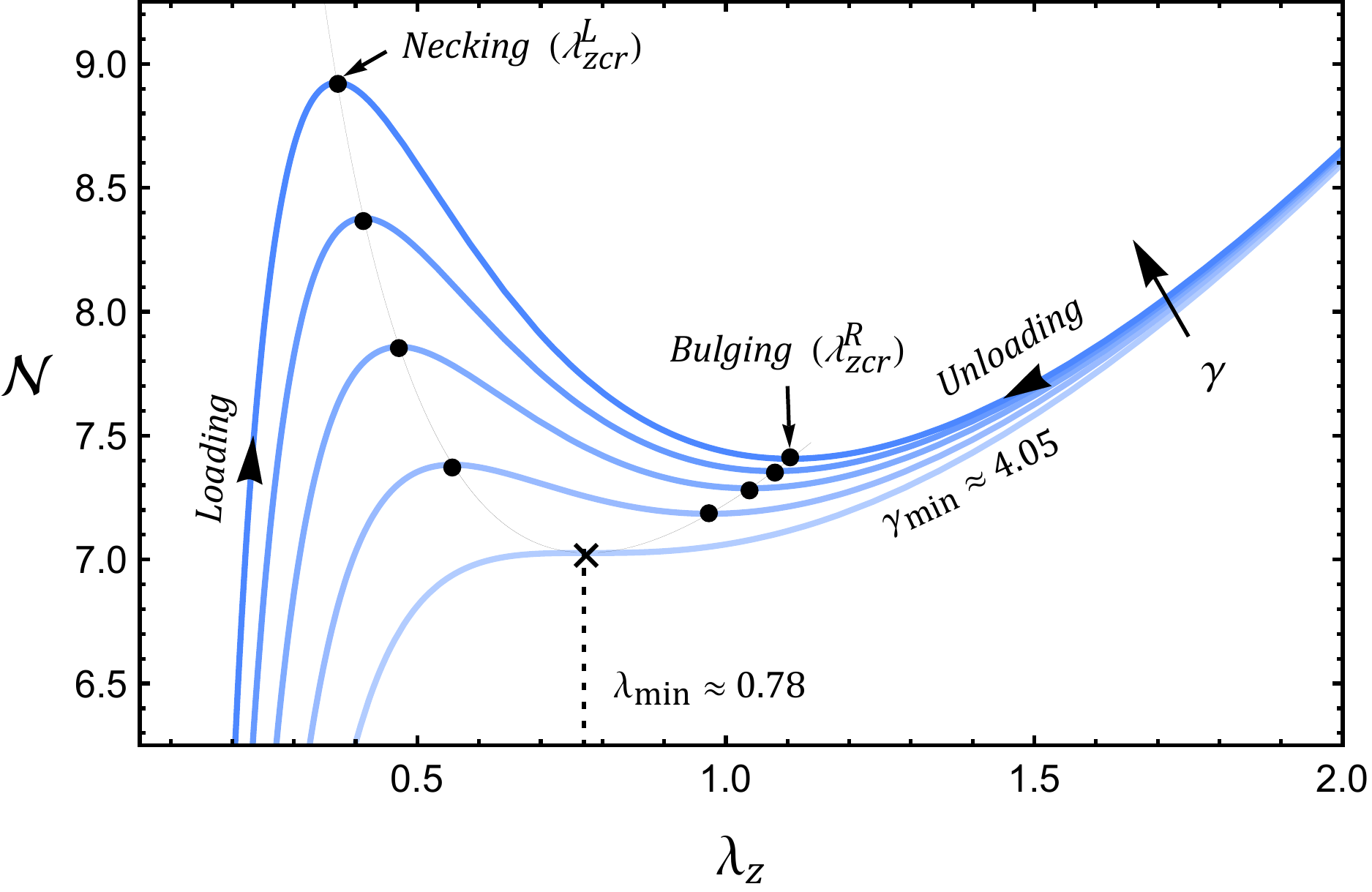}};
\node[text width=1cm] at (0.5,2.75)
    {\footnotesize \text{(a)}};
\node[text width=1cm] at (8.1,2.75)
    {\footnotesize \text{(b)}};
\end{tikzpicture}
\caption{Plots of $\mc{N}$ against $\lz$ (blue curves) for $\Jm=100$ and $\gamma=\gamma_{\tx{min}},4.5,5,5.5,6$, with (a) $\nu=0.05$ and (b) $\nu=0.25$. The bifurcation criterion $\mc{N}_{\tx{cr}}=\mc{N}(\lzcr)$ is given by the black curves. The arrows labelled ``$\gamma$" give the direction in which the fixed surface tension values are increasing.}
\label{NLv}
\end{figure}

In Fig. $\ref{NLv}$, for the compressible Gent model $(\ref{gentC})$ with $\Jm =100$, we plot the resultant axial force $\mc{N}$ against $\lz$ for several fixed $\gamma\geq 0$ (blue curves), as well as the bifurcation criterion $\mc{N}_{\tx{cr}}=\mc{N}(\lzcr)$ (black curves), with (a) $\nu=0.05$ and (b) $\nu=0.25$. We observe that, as in the incompressible case, there exists a minimum value of $\gamma$, $\gamma_{\tx{min}}$, below which the $\mc{N}=\mc{N}(\lz)$ curve is monotonic increasing and localization cannot occur. Here, the value of $\gamma_{\tx{min}}$ is dependent on the value of $\nu$. For any fixed $\gamma>\gamma_{\tx{min}}$, there exists two bifurcation values of $\lz$, $\lzcr^{L}$ and $\lzcr^{R}>\lzcr^{L}$, which are located at the local maximum and minimum of the now non-monotonic $\mc{N}=\mc{N}(\lz)$ curve. In the limit $\gamma\RA\gamma_{\tx{min}}$, these two bifurcation values coalesce into a single value $\lz=\lambda_{\tx{min}}$, and the maximum and minimum of $\mc{N}$ coalesce into an inflection point (marked by the black cross). When ``loading" from $\mc{N}=0$, the local maximum of $\mc{N}$ is the bifurcation point of interest, and we expect from the weakly non-linear analysis of \cite{FuST} and \cite{emery2021PRSA} that \textit{localized necking} will be triggered when this point is reached. In contrast, when ``unloading" from some large strictly positive $\mc{N}$, the local minimum of $\mc{N}$ is the relevant bifurcation point, and we expect that \textit{localized bulging} will initiate here. 

\begin{figure}[h!]
\centering
\begin{tikzpicture}
\node at (0,0) {\includegraphics[scale=0.378]{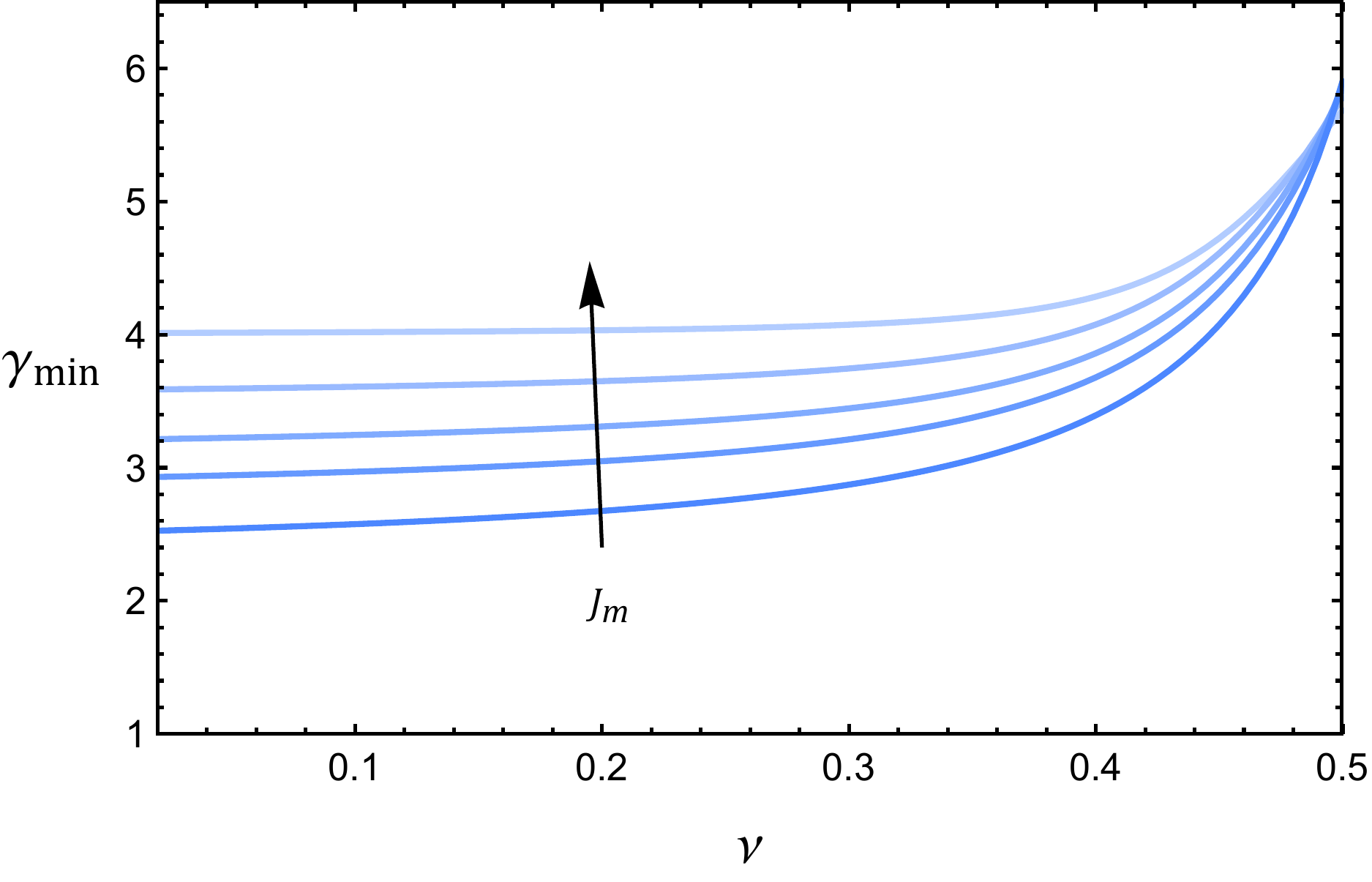}};
\node at (7.5,0) {\includegraphics[scale=0.38]{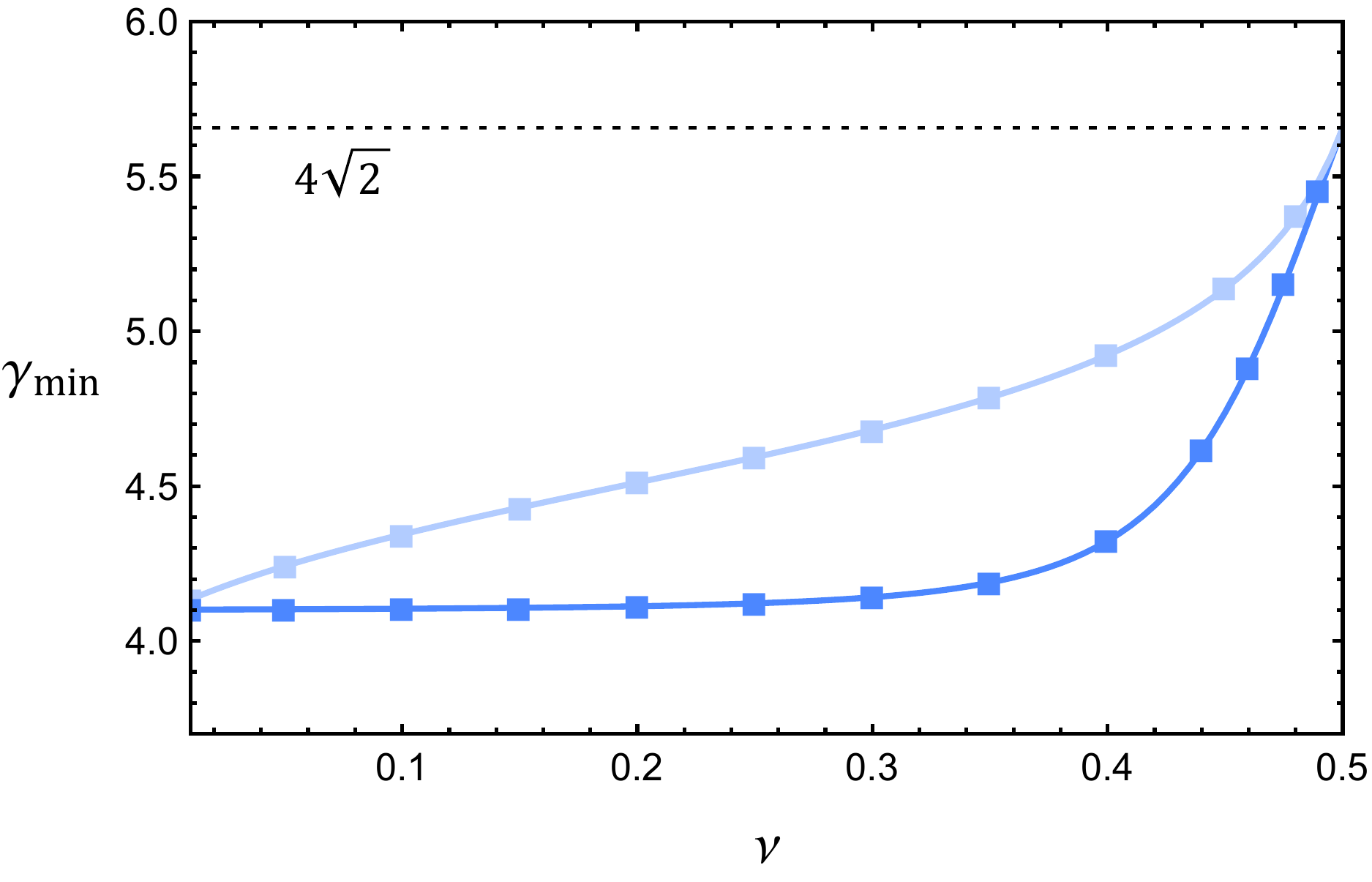}};
\node[text width=1cm] at (0.5,2.75)
    {\footnotesize \text{(a)}};
\node[text width=1cm] at (8.1,2.75)
    {\footnotesize \text{(b)}};
\end{tikzpicture}
\caption{Plots of $\gamma_{\tx{min}}$ against $\nu$ for (a) the Gent material model $(\ref{gentC})$ with $\Jm=3,5,7.5,15,100$ and (b) the quadratic (dark blue) and logarithmic (light blue) neo-Hookean models. In (b), the solid lines represent our theoretical results, and the squares give numerical results from DJ.}
\label{lgmin}
\end{figure}

We observe in Fig. $\ref{NLv}$ that a larger value of $\gamma>\gamma_{\tx{min}}$ will delay the expected onset of localized necking when ``loading" from $\mc{N}=0$, but incite the expected onset of localized bulging when ``unloading" from some large strictly positive $\mc{N}$. In Fig. $\ref{lgmin}$ (a), we examine the variation of $\gamma_{\tx{min}}$ with respect to $\nu$ for the Gent material model $(\ref{gentC})$ with several fixed values of $\Jm$. We observe that $\gamma_{\tx{min}}$ increases with both $\nu$ and $\Jm$. Thus, for materials with a greater level of compressibility (i.e. for values of $\nu$ closer to zero), or a lower level of extensibility, there is a greater range of values of $\gamma$ for which localization can occur in this loading scenario. In (b), we plot this same relationship for the quadratic (solid dark blue curve) and logarithmic (solid light blue curve) neo-Hookean material models, and compare with the numerical simulation results presented in Fig. $6$ (a) of DJ (squares). We note the exceptional agreement between both sets of results, and in the incompressible limit $\nu\RA 1/2$, we recover the value $\gamma_{\tx{min}}=4\sqrt{2}$ which was reported in \cite{FuST}.

\subsection{Fixed $\lz$ and increasing $\gamma$}
When $\lz\geq 1$ is fixed and $\gamma$ is increased gradually from zero, the bifurcation condition for localization is also $d\mc{N}/d\lz =0$. Hence, it is equivalent to $(\ref{bccompr})$ except the left hand side is replaced with $\gamma=\gcr$ here and the right hand side is evaluated at the fixed axial stretch rather than $\lzcr$.

\begin{figure}[h!]
\centering
\begin{tikzpicture}
\node at (0,0) {\includegraphics[scale=0.36]{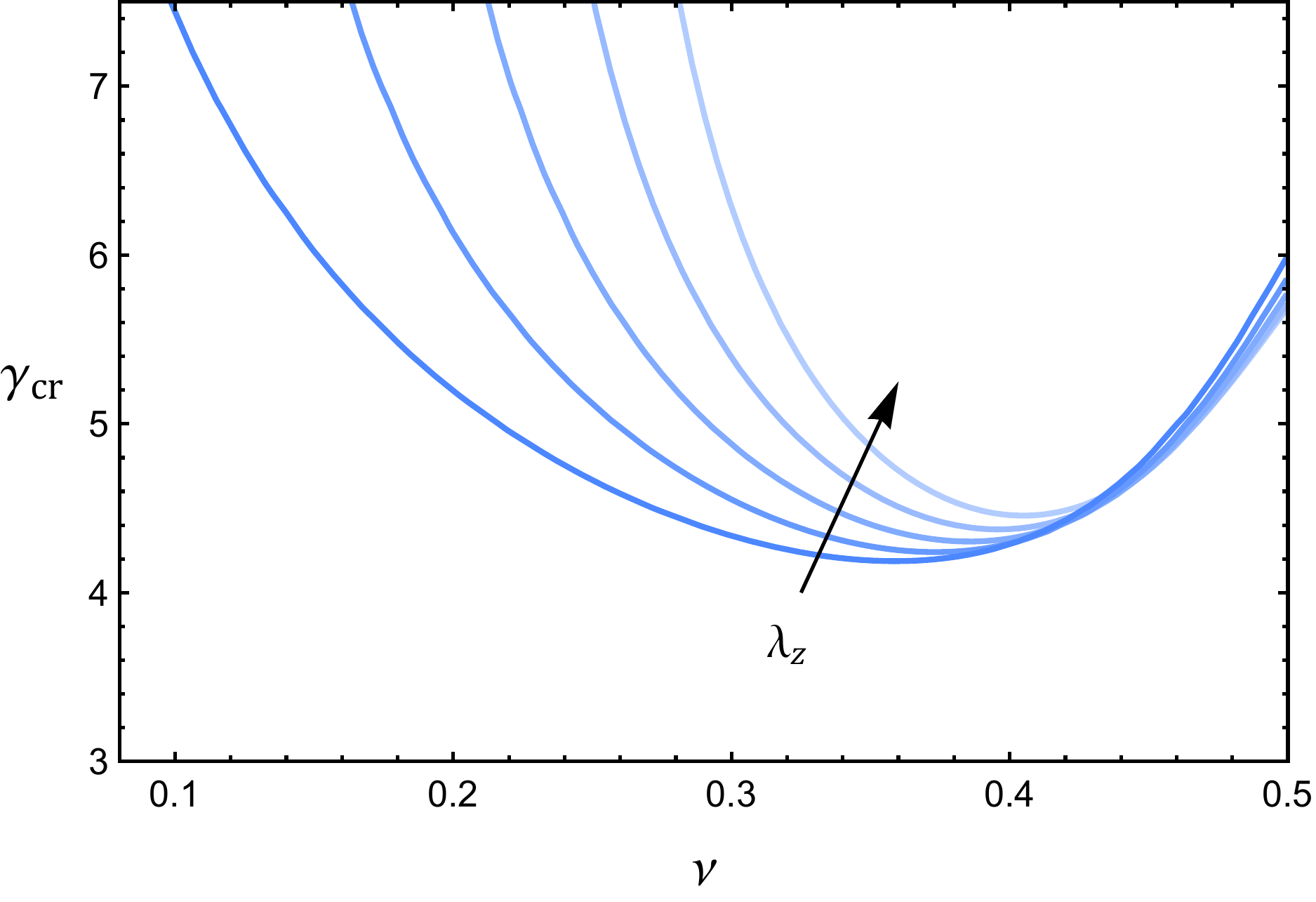}};
\node at (7.5,0) {\includegraphics[scale=0.37]{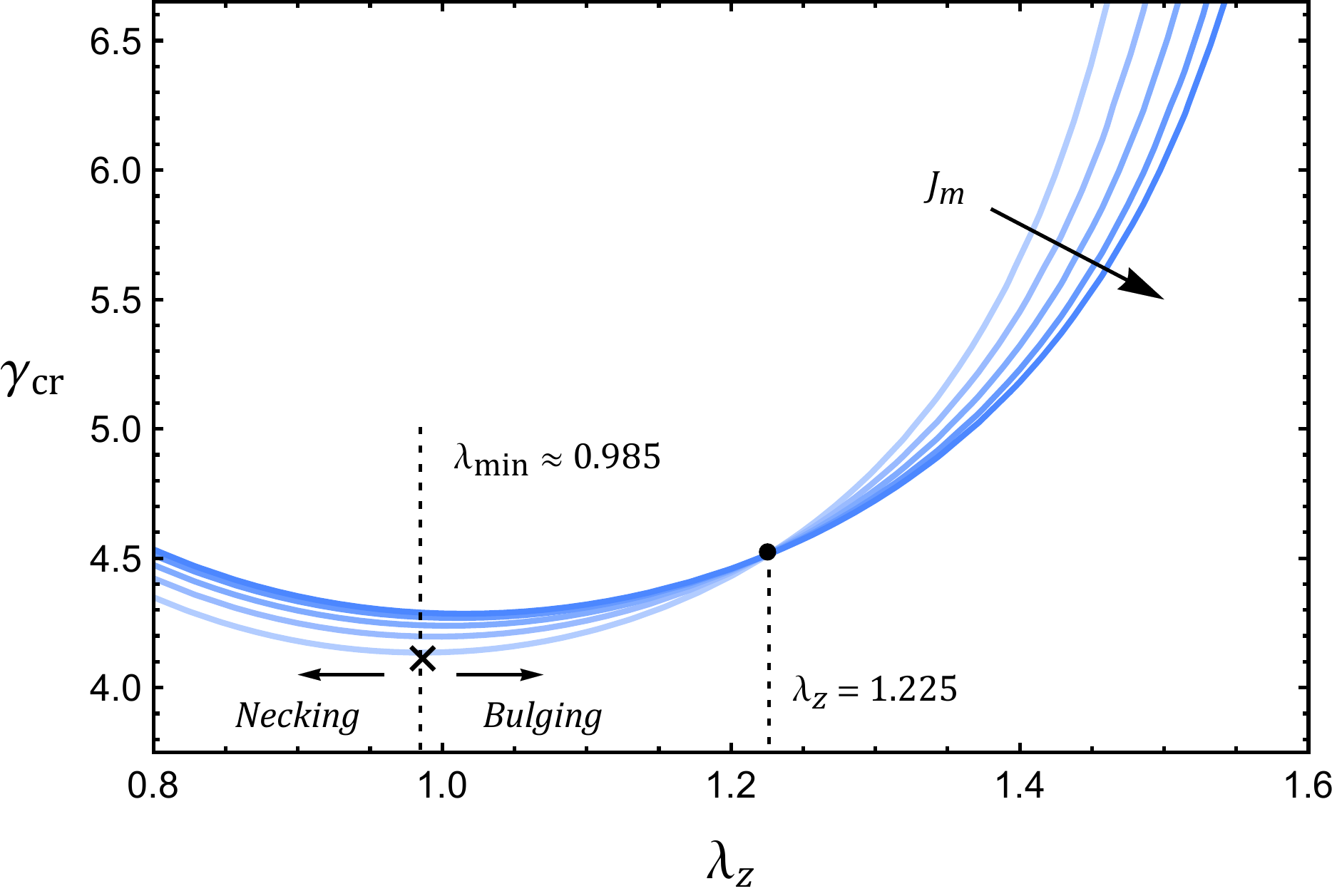}};
\node[text width=1cm] at (0.5,2.75)
    {\footnotesize \text{(a)}};
\node[text width=1cm] at (8.1,2.75)
    {\footnotesize \text{(b)}};
\end{tikzpicture}
\caption{Plots of $\gcr$ against (a) $\nu$ and (b) $\lz$ corresponding to the Gent material model $(\ref{gentC})$. In (a), we have $\Jm=100$ and $\lz=1,1.05,1.1,1.15,1.2$, and in (b) we have $\nu=0.4$ and $\Jm=20,30,45,70,100$. Arrows indicate the direction of parameter growth.}
\label{lgcrv}
\end{figure}

\begin{figure}[h!]
\centering
\begin{tikzpicture}
\node at (0,0) {\includegraphics[scale=0.365]{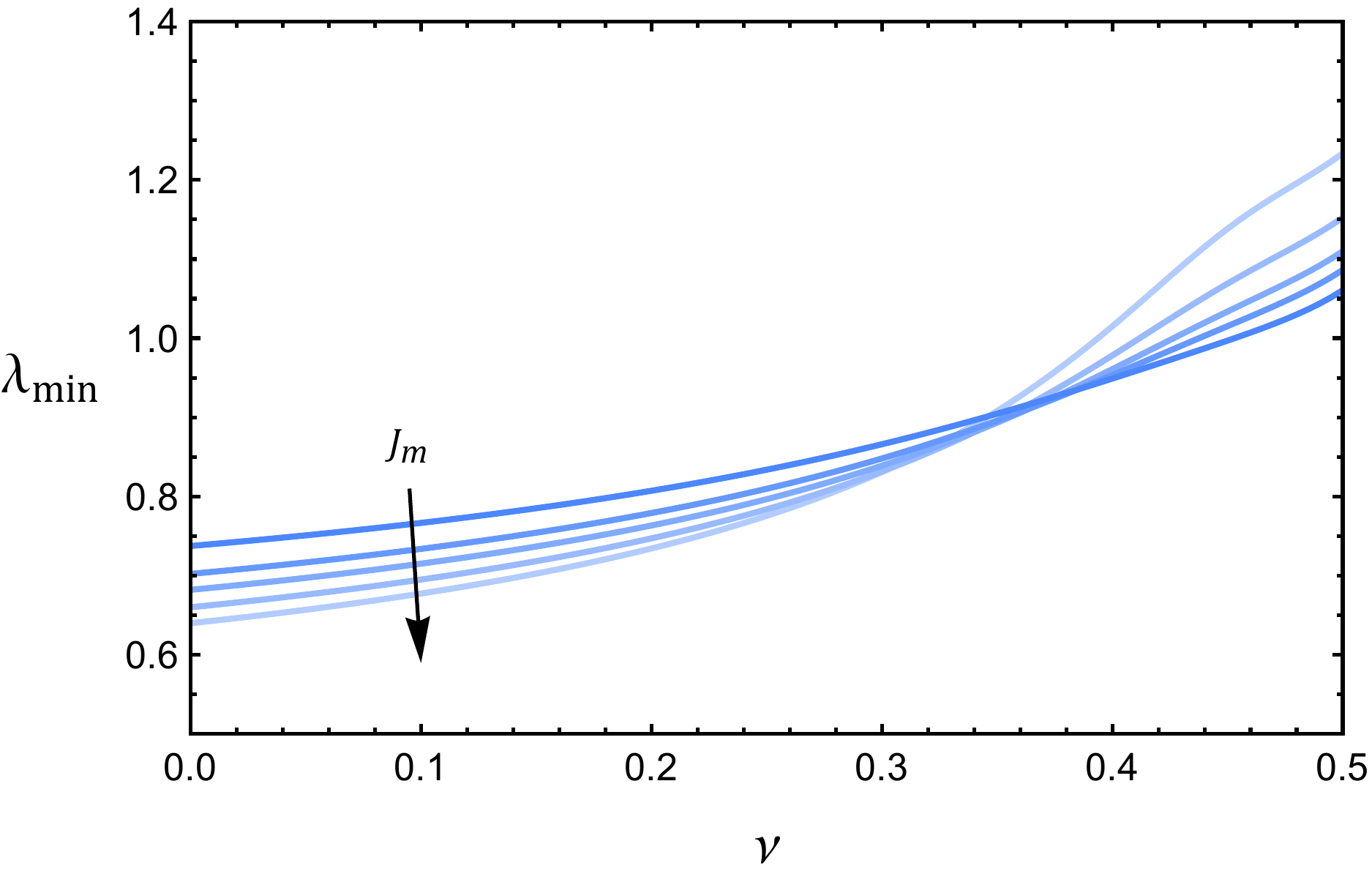}};
\node at (7.5,0) {\includegraphics[scale=0.37]{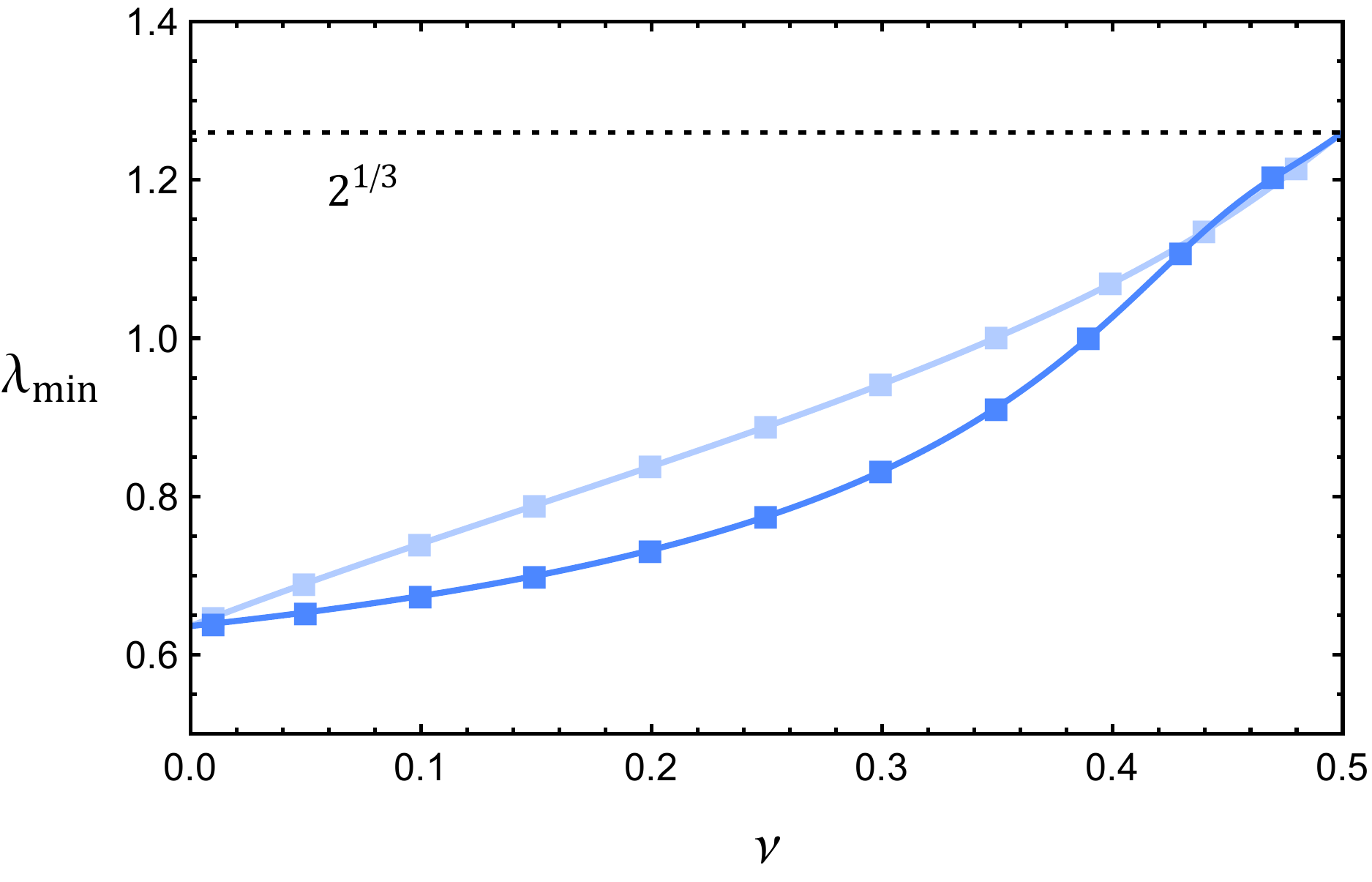}};
\node[text width=1cm] at (0.5,2.75)
    {\footnotesize \text{(a)}};
\node[text width=1cm] at (8.1,2.75)
    {\footnotesize \text{(b)}};
\end{tikzpicture}
\caption{Plots of $\lambda_{\tx{min}}$ against $\nu$ corresponding to (a) the Gent material model $(\ref{gentC})$ and (b) the quadratic (dark blue) and logarithmic (light blue) neo-Hookean material models. In (a), we fix $\Jm=3,5,7.5,15,100$, and the arrow indicates the direction of parameter growth. In (b), the solid curves give our theoretical results, and the squares give the numerical results of DJ. In the incompressible limit $\nu\RA 1/2$, we recover the value $\lambda_{\tx{min}}\RA 2^{1/3}$ reported in \cite{FuST}.}
\label{lminnuNHG}
\end{figure}

\begin{figure}[h!]
\centering
\includegraphics[scale=0.42]{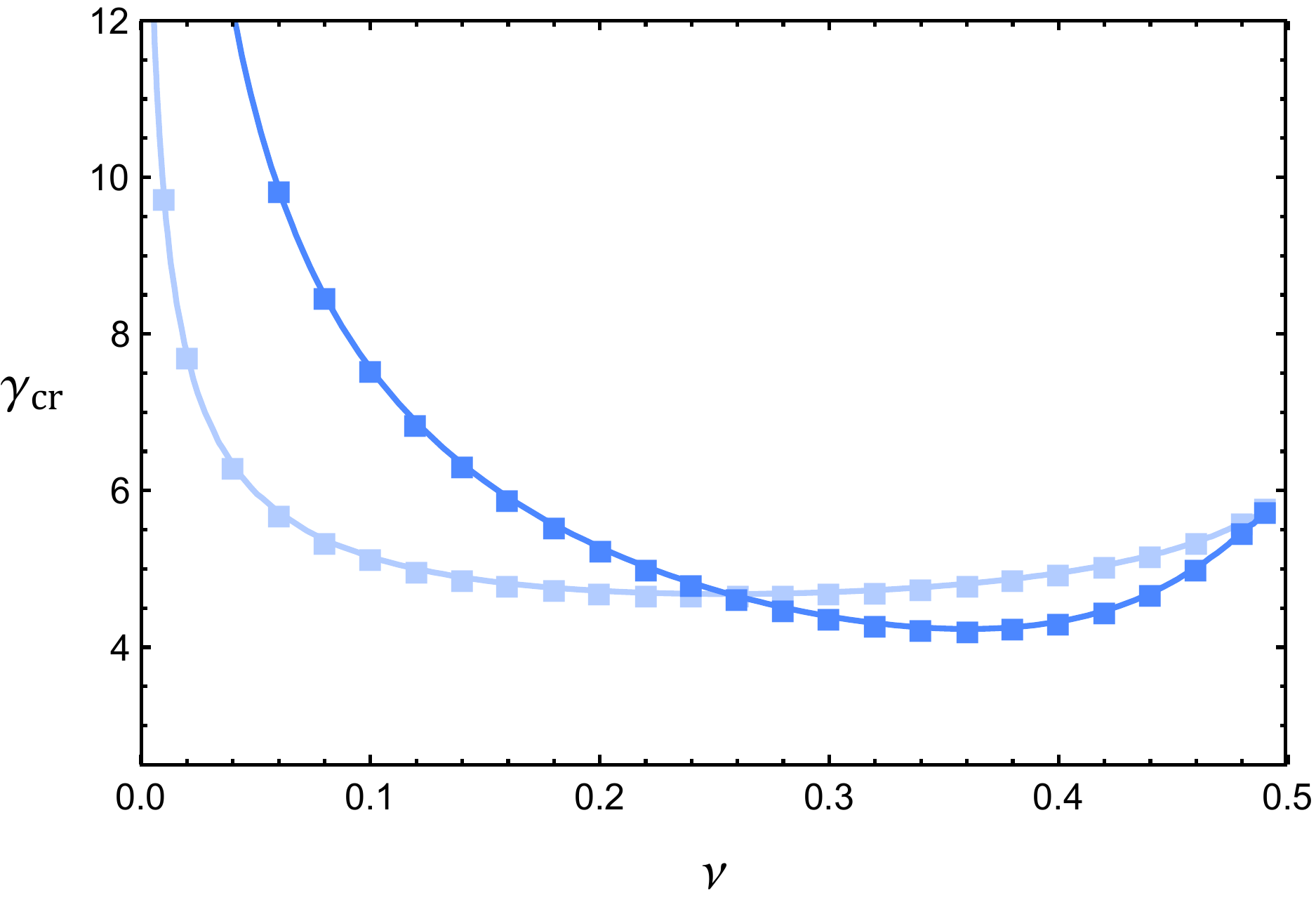}
\caption{A plot of $\gcr$ against $\nu$ for the quadratic (dark blue) and logarithmic (light blue) compressible neo-Hookean models. The solid curves give our theoretical results, whilst the squares give numerical results from DJ.}
\label{gcrnuNH}
\end{figure}

In Fig. $\ref{lgcrv}$, we plot $\gcr$ against (a) $\nu$ for $\Jm=100$ and several fixed values of $\lz\geq 1$ and (b) $\lz$ for $\nu=0.4$ and several fixed values of $\Jm$. In (a), we observe that highly compressible cylinders are the least susceptible to localization. We also observe that there exists a threshold value of $\nu$ below which a greater fixed axial stretch suppresses localization, and above which a greater fixed axial stretch encourages localization. In (b), we show that there exists a threshold value of $\lz$ ($\lz=1.225$ in the case presented) below which a greater extensibility limit suppresses localization, and above which a greater extensibility limit encourages localization. We observe that $\gcr$ as a function of $\lz$ also possesses a minimum. This behaviour was similarly observed in the incompressible case studied by \cite{FuST}, and it was shown that the initial bifurcation solution was localized necking (localized bulging) for $\lz<\lambda_{\text{min}}$ ($\lz>\lambda_{\text{min}}$). Here, the value of $\lambda_{\text{min}}$ will vary with $\nu$, and we plot this relationship in Fig. $\ref{lminnuNHG}$. We see that, for all values of $\Jm$ considered, the value of $\lambda_{\tx{min}}$ increases with $\nu$. Thus, for cylinders with a greater degree of compressibility, there is a larger range of values of fixed $\lz$ for which an initial localized bulging solution will emerge at $\gamma=\gcr$. In Fig. $5$ (b), we demonstrate the exceptional agreement between our theoretical results and the numerical simulation results in Fig. $6$ (b) of DJ for the quadratic and logarithmic neo-Hookean models.   
In Fig. $\ref{gcrnuNH}$, we plot the variation of $\gcr$ with respect to $\nu$ for the quadratic and logarithmic neo-Hookean models with $\lz=1$, and we note again that there is exceptional agreement between our theoretical results (solid curves) and the numerical simulation results given in Fig. 5 of DJ (squares).

\subsection{Fixed $\mc{N}$ and increasing $\gamma$}

By making appropriate rearrangements in $(\ref{compN})$, the surface tension $\gamma$ can be expressed explicitly in terms of $\lambda_{\theta}$ and $\lz$ for any fixed $\mc{N}\geq 0$ as follows:
\begin{align}
\gamma = \frac{1}{2}\left\{\frac{\mc{N}}{\pi\lambda_{\theta}\Ro} - 2\Ro\lambda_{\theta}^{3}\lz W_{3} - 2\Ro\lambda_{\theta}^{-1}W_{1}\right\}. \label{gITON}
\end{align}
Then, we may subtract $(\ref{gITON})$ from $(\ref{compgamrel})$ and define $\lambda_{\theta}$ as an implicit function of $\lz$ from the resulting equation. The bifurcation condition for localization is then conjectured to be $d\gamma/d\lz =0$, where $\gamma$ is given in $(\ref{gITON})$ and $\mc{N}$ is fixed in the differentiation. Explicitly, this condition is expressible as
\begin{align}
\frac{\lambda_{\theta\text{d}}\mc{N}}{2\pi\Ro^2}&=\left\{W_{1}-\lambda_{\theta}^{2}\left(4W_{11}+3\lambda_{\theta}^{2}\lz W_{3}+4\lambda_{\theta}^{2}\lz\left((\lz + \lambda_{\theta}^2)W_{13}+\lambda_{\theta}^{4}\lz^{2}W_{33}\right)\right)\right\}\lambda_{\theta\text{d}}\nonumber\\[0.4em]
&\,\,\,\,\,\,\,\,-2\lambda_{\theta}\lz W_{11}-\lambda_{\theta}^{5}\left\{W_{3}+2\lz\left((\lz +1)W_{13} + \lambda_{\theta}^{4}\lz W_{33}\right)\right\}, \label{compBCN}
\end{align}
and this equation is also evaluated at $\lz=\lzcr$. The expression for $\lambda_{\theta\text{d}}$ can again be obtained by differentiating $(\ref{compgamrel})$ implicitly with respect to $\lz$. The left hand side of the resulting equation will vanish since $d\gamma/d\lz=0$ is the bifurcation condition in this loading scenario, and we thus recover the relation presented in $(\ref{ltdd})$. Once the critical stretch $\lzcr$ has been obtained from $(\ref{compBCN})$, we can substitute this value into the equation $(\ref{gITON})$ to obtain the corresponding critical surface tension $\gamma_{\tx{cr}}\equiv\gamma(\lzcr)$.

\begin{figure}[h!]
\centering
\begin{tikzpicture}
\node at (0,0) {\includegraphics[scale=0.365]{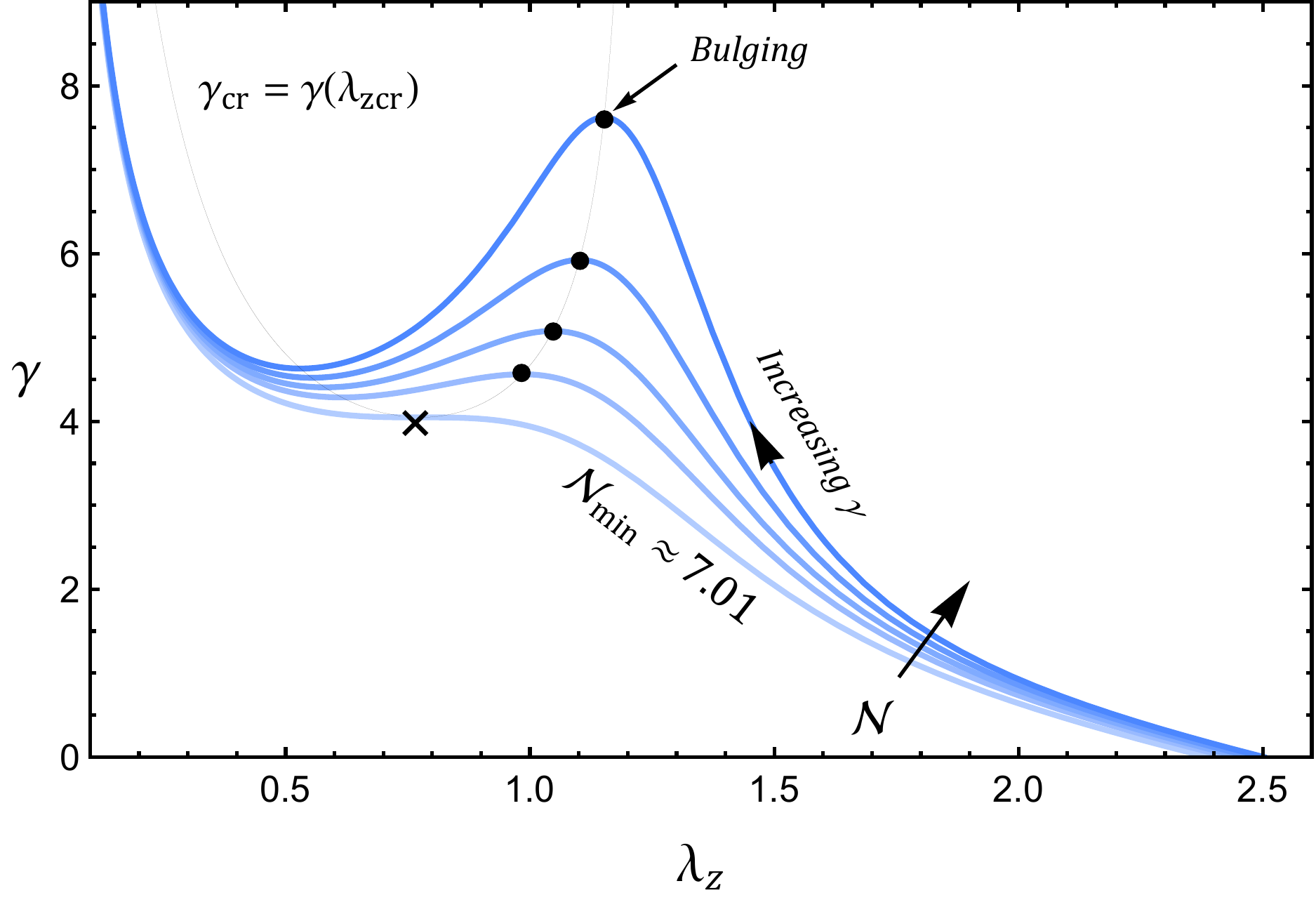}};
\node at (7.5,0) {\includegraphics[scale=0.37]{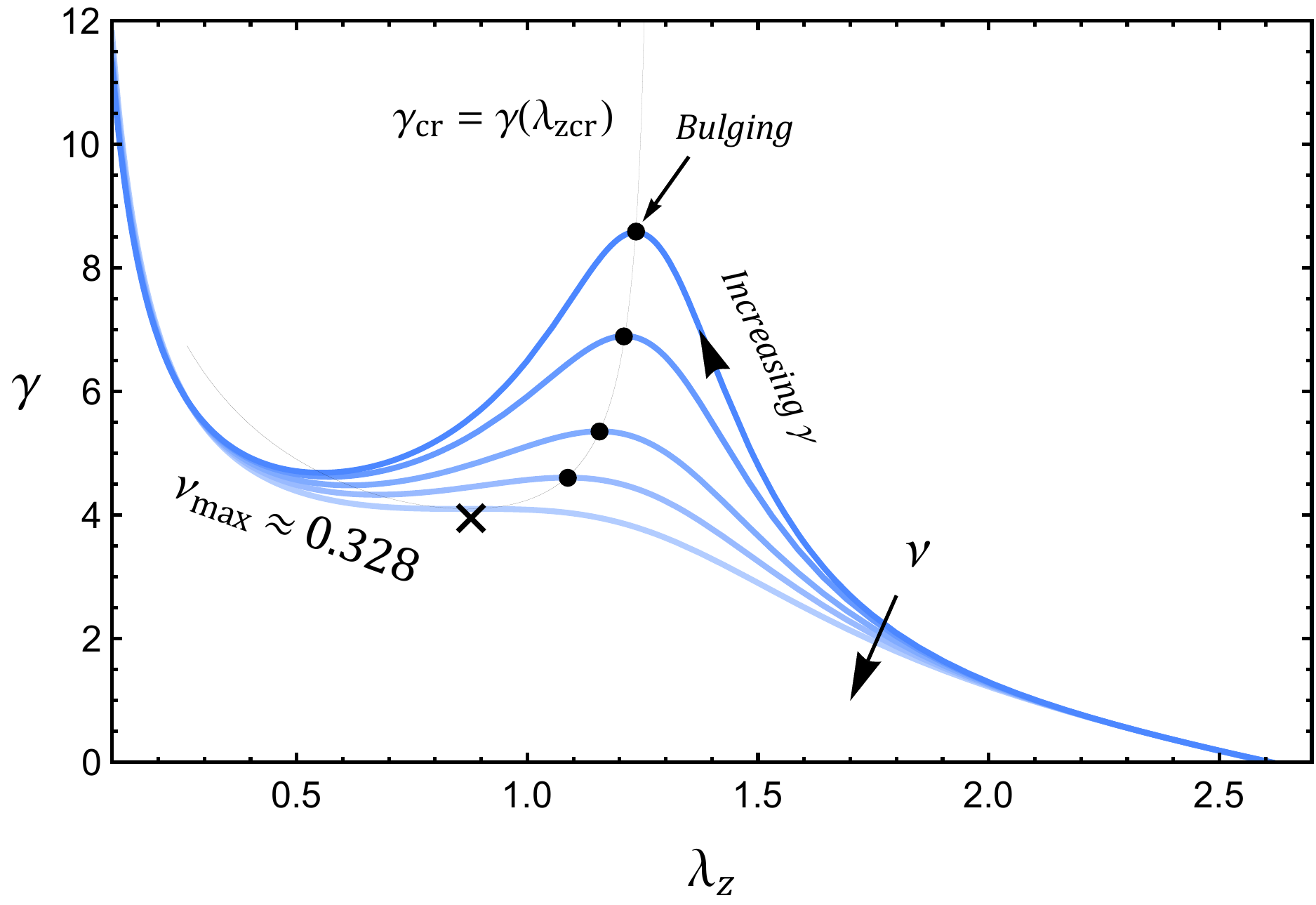}};
\node[text width=1cm] at (0.5,2.75)
    {\footnotesize \text{(a)}};
\node[text width=1cm] at (8.1,2.75)
    {\footnotesize \text{(b)}};
\end{tikzpicture}
\caption{Plots of $\gamma$ against $\lz$ (blue curves) corresponding to the Gent material model $(\ref{gentC})$ with $\Jm=100$. In (a) we fix $\nu=0.25$ and $\mc{N}=\mc{N}_{\tx{min}},7.2,7.3,7.4,7.5$, and in (b) we fix $\mc{N}=8$ and $\nu=0.29,0.295,0.305,0.315,\nu_{\tx{max}}$, where $\mc{N}_{\tx{min}}\approx 7.01$ and $\nu_{\tx{max}}\approx 0.328$. The black curves give the bifurcation criterion $\gamma_{\tx{cr}}\equiv\gamma(\lzcr)$, and arrows indicate the direction of parameter growth.}
\label{glamNn}
\end{figure}

In Fig. $\ref{glamNn}$, we plot the function $\gamma=\gamma(\lz)$ given in $(\ref{gITON})$ for (a) $\nu=0.25$, $\Jm=100$ and several fixed $\mc{N}\geq 0$, and (b) $\mc{N}=8$, $\Jm=100$ and several fixed $\nu$. As in the purely incompressible case, there is seen in (a) to be a minimum value of $\mc{N}$, $\mc{N}_{\tx{min}}$, below which the curve $\gamma=\gamma(\lambda)$ is monotonic decreasing and localization is prohibited. This minimum value of $\mc{N}$ will depend on the value of $\nu$. Above this minimum value of $\mc{N}$, two bifurcation values of $\lz$, $\lzcr^{L}$ and $\lzcr^{R}>\lzcr^{L}$, emerge and correspond to the local minimum and maximum of the now non-monotonic $\gamma=\gamma(\lz)$ curve, respectively. When fixing $\mc{N}>\mc{N}_{\tx{min}}$ with $\gamma=0$ initially, an initial axial stretch $\lz>\lzcr^{R}$ will be produced. As we then increase $\gamma$ monotonically from zero, we will approach $\lz=\lzcr^{R}$ from the right (recall that an increase in $\gamma$ must be accompanied  by a decrease in $\lz$ to preserve the constant $\mc{N}$). Thus, the bifurcation point of interest in this loading scenario is situated at the local maximum of $\gamma=\gamma(\lz)$, and we expect that a \textit{localized bulge} will initiate at this point given the findings of \cite{FuST} for the incompressible case. 

In Fig. $\ref{glamNn}$ (a), we observe that larger fixed $\mc{N}$ above $\mc{N}_{\tx{min}}$ correspond to larger values of $\gcr$ (marked by the black dots), and so a greater fixed axial force will discourage localized bulging when the material is compressible. In (b),  we find that there exists a maximum value of $\nu$, $\nu_{\tx{max}}$, above which the $\gamma=\gamma(\lambda)$ curve becomes monotonic decreasing and localization becomes impossible.  The value of $\nu_{\tx{max}}$ will vary with the value of the fixed $\mc{N}$. We observe also that, for smaller values of $\nu$ (i.e. for materials with a greater level of compressibility), the associated value of $\gcr$ is larger. Hence, increased compressibility discourages the initiation of a localized bulge in this loading scenario. 

\begin{figure}[h!]
\centering
\begin{tikzpicture}
\node at (0,0) {\includegraphics[scale=0.365]{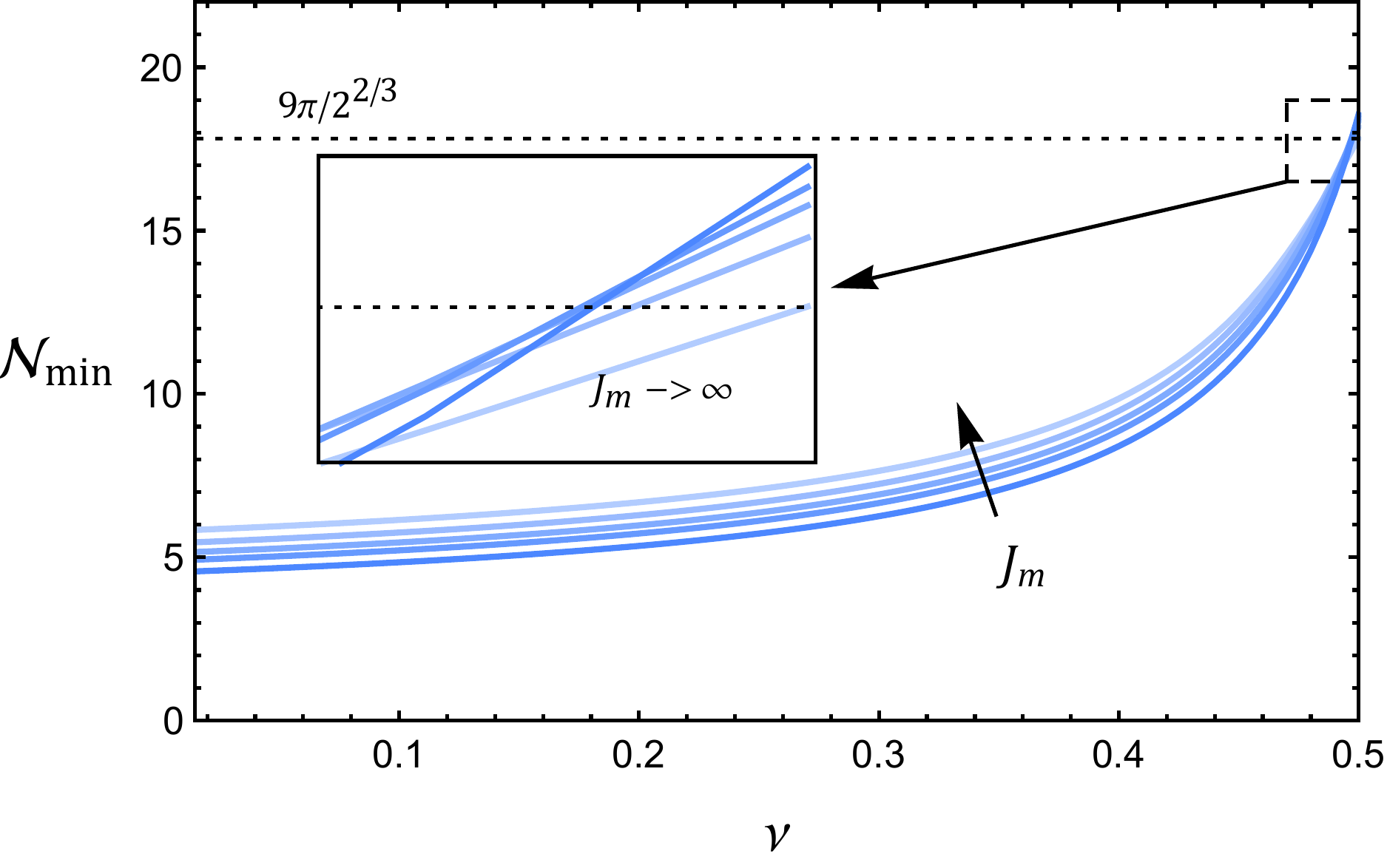}};
\node at (7.5,0) {\includegraphics[scale=0.36]{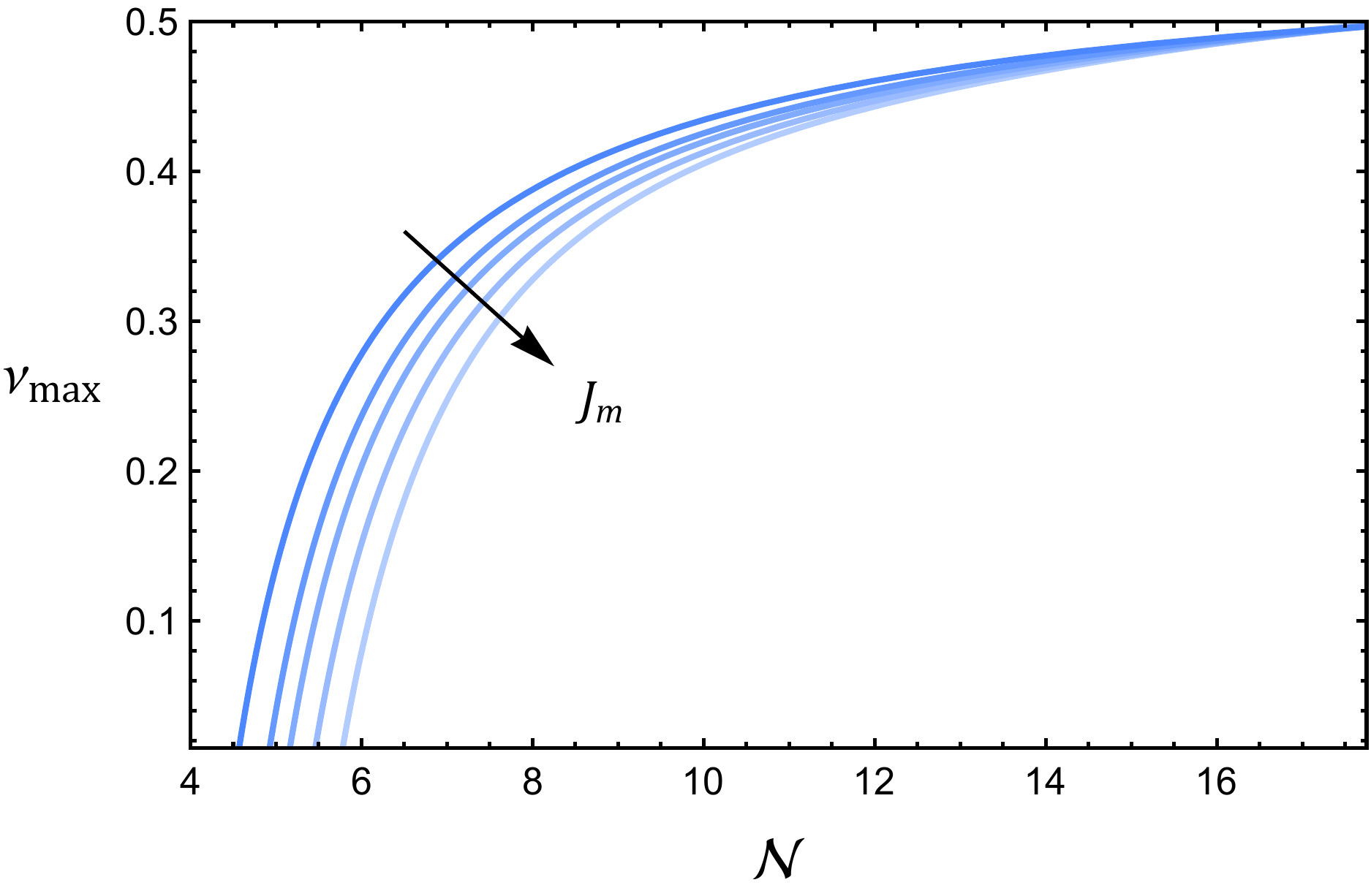}};
\node[text width=1cm] at (0.5,2.75)
    {\footnotesize \text{(a)}};
\node[text width=1cm] at (8.1,2.75)
    {\footnotesize \text{(b)}};
\end{tikzpicture}
\caption{The variation of \textbf{(a)} $\mc{N}_{\tx{min}}$ against $\nu$ and \textbf{(b)} $\nu_{\tx{max}}$ against $\mc{N}$ for the Gent material model $(\ref{gentC})$ with $\Jm=3,5,7.5,15$ and $\Jm\RA\infty$. Arrows indicate the direction of parameter growth.}
\label{Nvext}
\end{figure}

In Fig. $\ref{Nvext}$, we plot the variation of (a) $\mc{N}_{\tx{min}}$ against $\nu$ and (b) $\nu_{\tx{max}}$ against $\mc{N}$. In (a), for any given Poisson ratio $\nu$, localized bulging is only possible if the fixed axial force $\mc{N}$ is greater than the value of $\mc{N}_{\tx{min}}$ given by the appropriate blue curve. For all values of $\Jm$ considered, the value of $\mc{N}_{\tx{min}}$ increases with $\nu$. Thus, for materials with a greater degree of compressibility, there is a greater range of values of fixed $\mc{N}$ for which localized bulging can occur. We note also that, in the incompressible limit $\nu\RA 1/2$, we recover the result $\mc{N}_{\tx{min}}=9\pi/2^{2/3}$ in the limit $\Jm\RA\infty$ which was originally given in Fig. 5 (b) of \cite{FuST}. In (b), for any given fixed value of $\mc{N}$, localized bulging is only possible provided the value of the Poisson's ratio is less than the value $\nu_{\tx{max}}$ on the appropriate blue curve shown. We observe that, for each value of $\Jm$, there exists a value of $\mc{N}$ below which localized bulging is prohibited in cylinders with \textit{any} degree of compressibility. For instance, in the limit $\Jm\RA\infty$, localized bulging is impossible in any compressible cylinder if  $\mc{N}< 5.825$.


\section{Post-bifurcation behaviour}
Our analysis in the previous section provided a generalized analytical framework to support the numerical simulation predictions of DJ for the bifurcation points corresponding to elasto-capillary localization in compressible solid cylinders. In this section, we go one step further by showing that the post-bifurcation behaviour of the cylinder can be comprehensively described with the aid of our analytical expression $(\ref{compN})$ for the primary deformation. The importance of such an analytical approach in corroborating post-bifurcation results from numerical simulations is also highlighted.

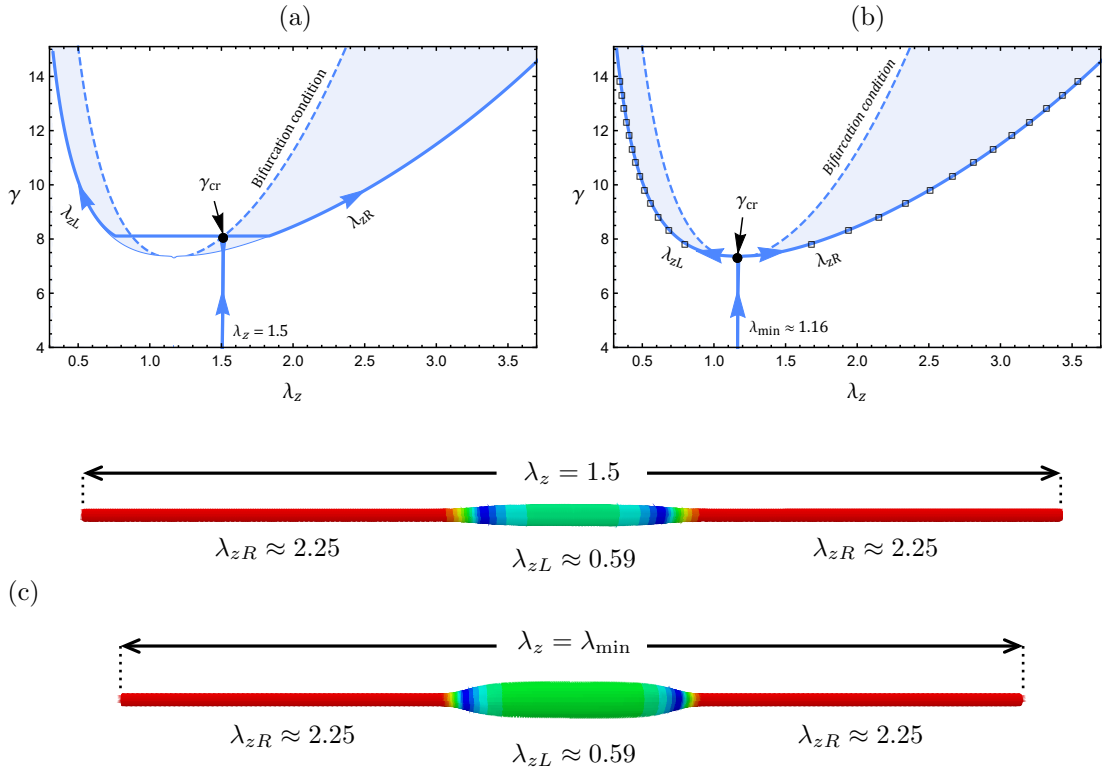
\begin{figure}[h!]
\centering
\begin{tikzpicture}
\node at (0,2.65) {\includegraphics[scale=0.4]{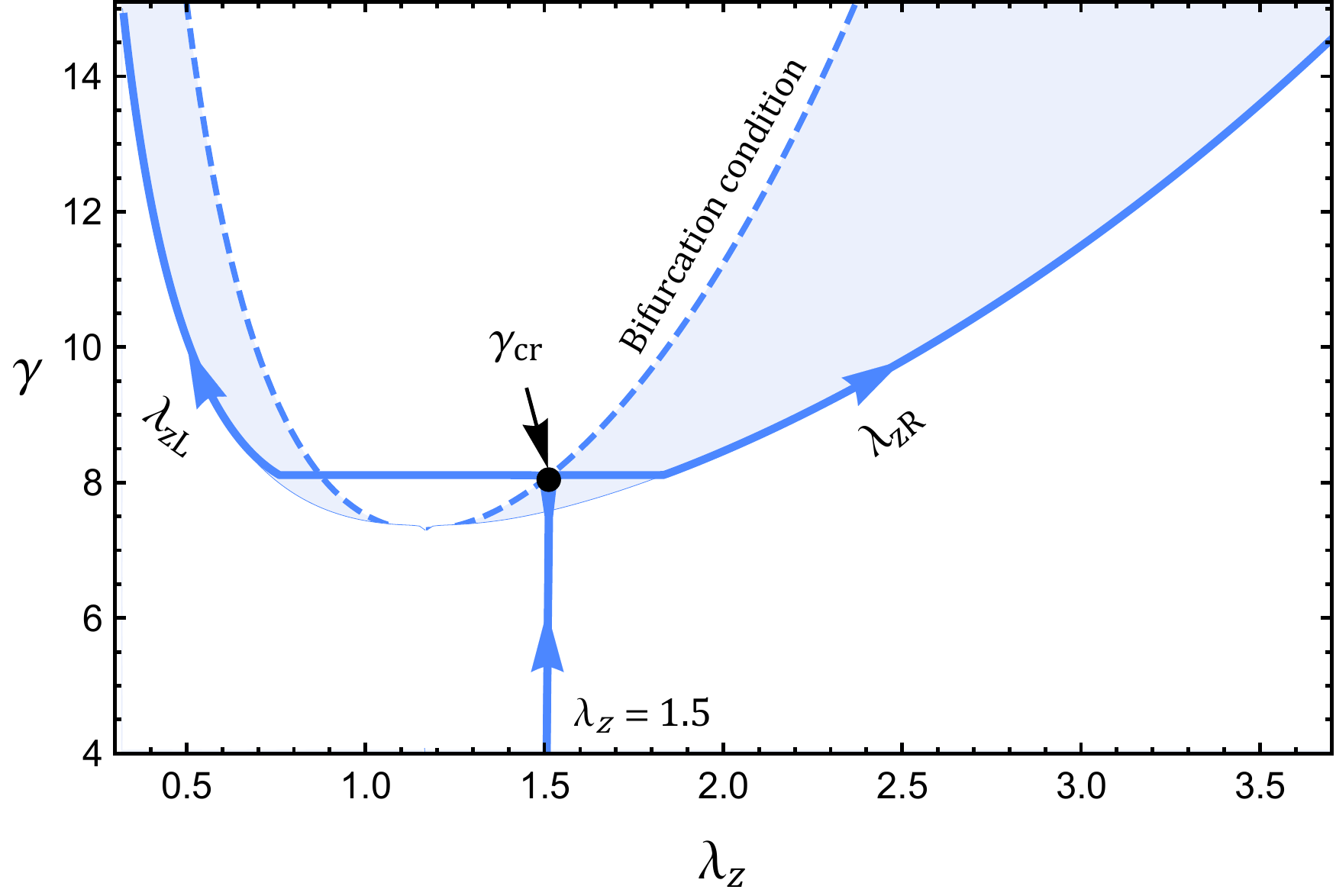}};
\node at (7.5,2.65) {\includegraphics[scale=0.4]{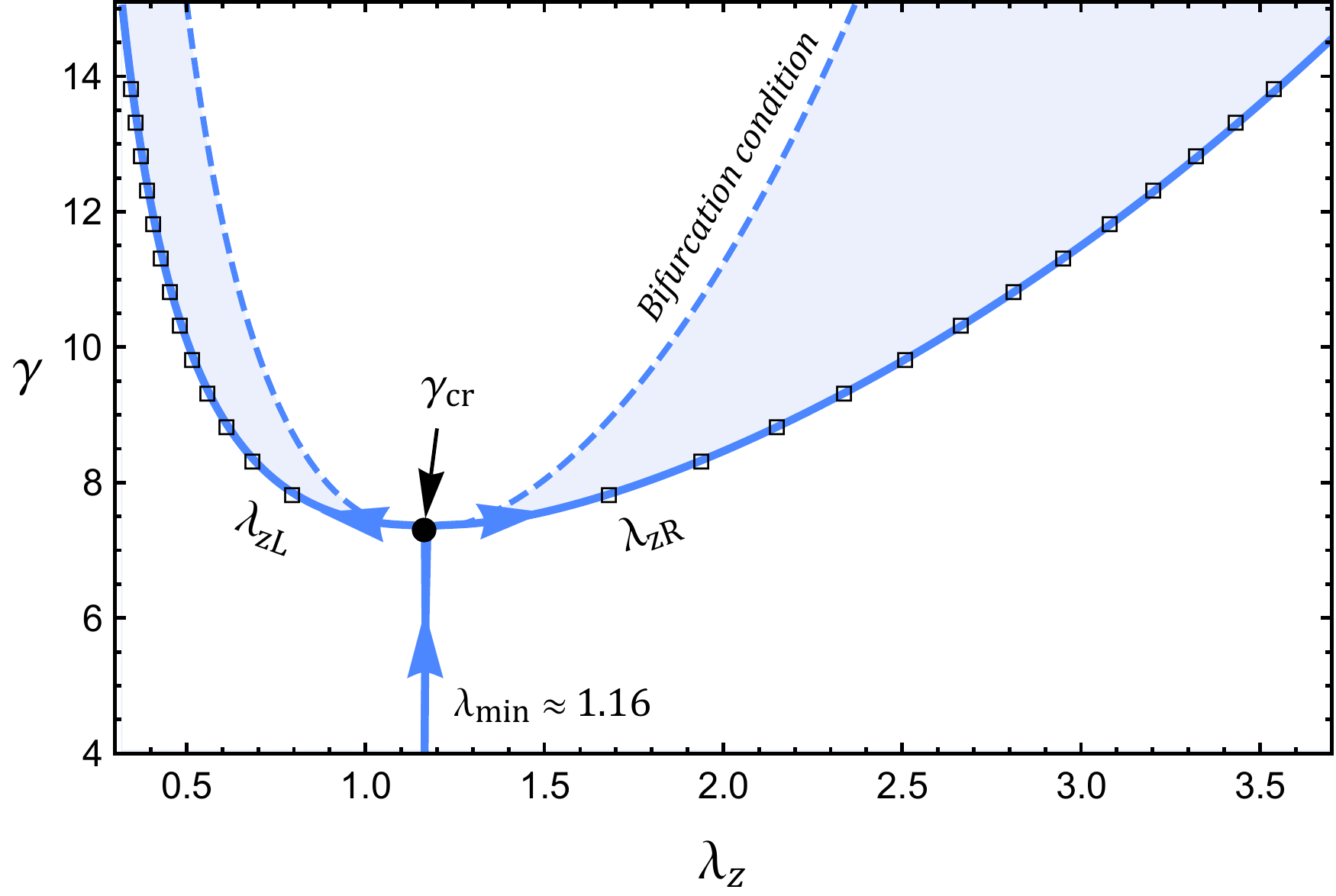}};
\node at (4,-1.25) {\includegraphics[scale=0.42]{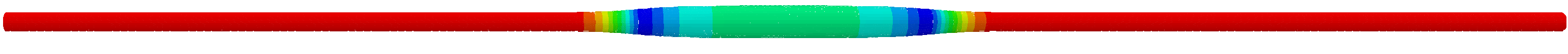}};
\node at (4,-3.65) {\includegraphics[scale=0.4]{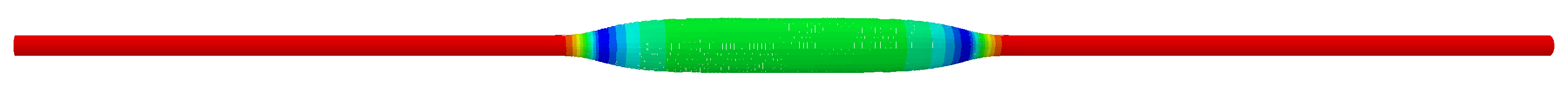}};
\node[text width=1cm] at (0.6,5.4)
    {\footnotesize \text{(a)}};
\node[text width=1cm] at (8.2,5.4)
    {\footnotesize \text{(b)}};
\node[text width=1cm] at (-3,-2.25)
    {\footnotesize \text{(c)}};
\node[text width=2cm] at (0.45,-4.15)
    {\footnotesize $\lambda_{zR}\approx 2.25$};
\node[text width=2cm] at (8,-4.15)
    {\footnotesize $\lambda_{zR}\approx 2.25$};
\node[text width=2cm] at (4.2,-4.4)
    {\footnotesize $\lambda_{zL}\approx 0.59$};
\node[text width=2cm] at (4.2,-1.8)
    {\footnotesize $\lambda_{zL}\approx 0.59$};
\node[text width=2cm] at (8.2,-1.65)
    {\footnotesize $\lambda_{zR}\approx 2.25$};
\node[text width=2cm] at (0.25,-1.65)
    {\footnotesize $\lambda_{zR}\approx 2.25$};
\node[text width=2cm] at (4.275,-2.9)
    {\footnotesize $\lz = \lambda_{\tx{min}}$};
\node[text width=2cm] at (4.36,-0.65)
    {\footnotesize $\lz = 1.5$};
\draw[>=angle 45, ->, very thick] (3,-2.95) -- (-2,-2.95);
\draw[>=angle 45, ->, very thick] (5,-2.95) -- (10,-2.95);
\draw[>=angle 45, ->, very thick] (3,-0.65) -- (-2.5,-0.65);
\draw[>=angle 45, ->, very thick] (5,-0.65) -- (10.5,-0.65);
\draw [line width=0.35mm,dotted] (-2,-2.95)--(-2,-3.52);
\draw [line width=0.35mm,dotted] (10,-2.95)--(10,-3.52);
\draw [line width=0.35mm,dotted] (-2.5,-1.05)--(-2.5,-0.65);
\draw [line width=0.35mm,dotted] (10.5,-1.05)--(10.5,-0.65);
\end{tikzpicture}
\caption{FEM simulation results (solid blue curve) for the case of a hollow tube with fixed inner radius $\Ri=0.4$, $\Jm=100$, \text{(a)} fixed $\lambda=\lambda_{\tx{min}}$ and \text{(b)} fixed $\lambda=1.5$. The dashed blue curves represent the theoretical bifurcation condition, and the black squares give the relationship between the surface tension $\gamma$ and stretches $\lambda_{L}$ and $\lambda_{R}$ determined from $(\ref{MEAR})$. The black dots mark the bifurcation point in each case given by the simulations. In \text{(c)}, we present the ``two-phase" configuration of the tube for fixed $\lambda=1.5$ and $\lambda_{\tx{min}}$ when the surface tension has been increased beyond its bifurcation value to $\gamma=9$. Both configurations consist of a bulged section with uniform axial stretch $\lambda_{zL}\approx0.59$, in between two depressed sections with stretch $\lambda_{zR}\approx 2.25$. The proportion of the bulged ``phases" differ in each case due to the different averaged axial stretches.}
\label{FEM}
\end{figure}

We focus primarily on the fixed $\lz$ and increasing $\gamma$ scenario, since this is arguably the least well understood case. To set the scene, we first review what is already known for the case of an incompressible hollow tube where the radial displacement of the inner surface is zero. This case was well studied in \cite{emery2021IJSS,emery2021PRSA}, and we note that an incompressible solid cylinder is recovered when the fixed inner radius $\Ri$ tends to zero. Finite Element Method simulations conducted in \cite{ab2013} are presented in Fig. $\ref{FEM}$ (solid blue curves) for the incompressible Gent material model with $\Jm=100$, $\Ri/\Ro =0.4$, (a) $\lz=1.5>\lambda_{\tx{min}}$ and (b) $\lz=\lambda_{\tx{min}}\approx 1.16$. The initial bifurcation occurs when $d\mc{N}/d\lz =0$, and we present this condition through the dashed blue curves. Through a weakly non-linear analysis, the initial bifurcation solution for $\lz=1.5>\lambda_{\tx{min}}$ was determined in \cite{emery2021PRSA} to emerge \textit{sub-critically} and to take the explicit form of a localized bulge. The numerical simulations then showed that, if we attempt to increase $\gamma$ beyond its bifurcation value $\gcr$, a \textit{snap-through} to a ``two-phase" state occurs. This configuration consists of an axially propagated bulge ``phase" with constant stretch $\lambda_{zL}$ and a depressed ``phase" with axial stretch $\lambda_{zR}<\lambda_{zL}$; these ``phases" are connected by a smooth yet sharp transition zone as shown in Fig. $\ref{FEM}$ (c). Note that the bulged ``phase" could either be centred at $z=0$ or situated as the two ends of the tube (depending on how the small amplitude imperfection in the simulations is introduced), and that the overall averaged axial stretch of the ``two-phase" state remains fixed at $1.5$. The stretches $\lambda_{zL}$ and $\lambda_{zR}$ are given by the left and right branches of numerical simulation curves in (a) and (b). In the case shown in (b) where $\lz=\lambda_{\tx{min}}$, an exceptionally \textit{super-critical} bifurcation takes place when the bifurcation value $\gamma=\gamma_{\tx{min}}$ is reached in which the tube evolves smoothly into the same ``two-phase" state as in the $\lz=1.5$ case without any initial localization. The difference between the two cases covered is that the overall averaged axial stretches are different, meaning that the proportion $\lambda_{zL}/\lz$ of the propagated bulge ``phase" with respect to the overall fixed length of the tube will differ; see (c).

Numerical simulations aren't, however, necessary to determine $\lambda_{zL}$ and $\lambda_{zR}$; they can be determined with the aid of the analytical expression for $\mc{N}=\mc{N}(\lz)$. To elaborate, as was first elucidated by \cite{JCM1875}, the stretches for each ``phase" can be defined implicitly as functions of $\gamma$ through the following \textit{equal area rule}:
\begin{align}
\mc{N}_{MW}\equiv\mc{N}(\lambda_{zL})=\mc{N}(\lambda_{zR}),\spc\int_{\lambda_{zL}}^{\lambda_{zR}}\mc{N}d\lz = (\lambda_{zR} - \lambda_{zL})\mc{N}(\lambda_{zL}). \label{MEAR}
\end{align}
The resultant axial force $\mc{N}_{MW}$ for the ``two-phase" state is invariant across the bulged and depressed ``phase", but will differ for each $\gamma>\gcr$.  In the aforementioned hollow tube case, exceptional agreement has previously been shown between the values of the Maxwell stretches produced through FEM simulations and Maxwell's equal area rule; see \cite{emery2021IJSS} and Fig. $\ref{FEM}$ (b).

\begin{figure}[h!]
\centering
\begin{tikzpicture}
\node at (0,0) {\includegraphics[scale=0.365]{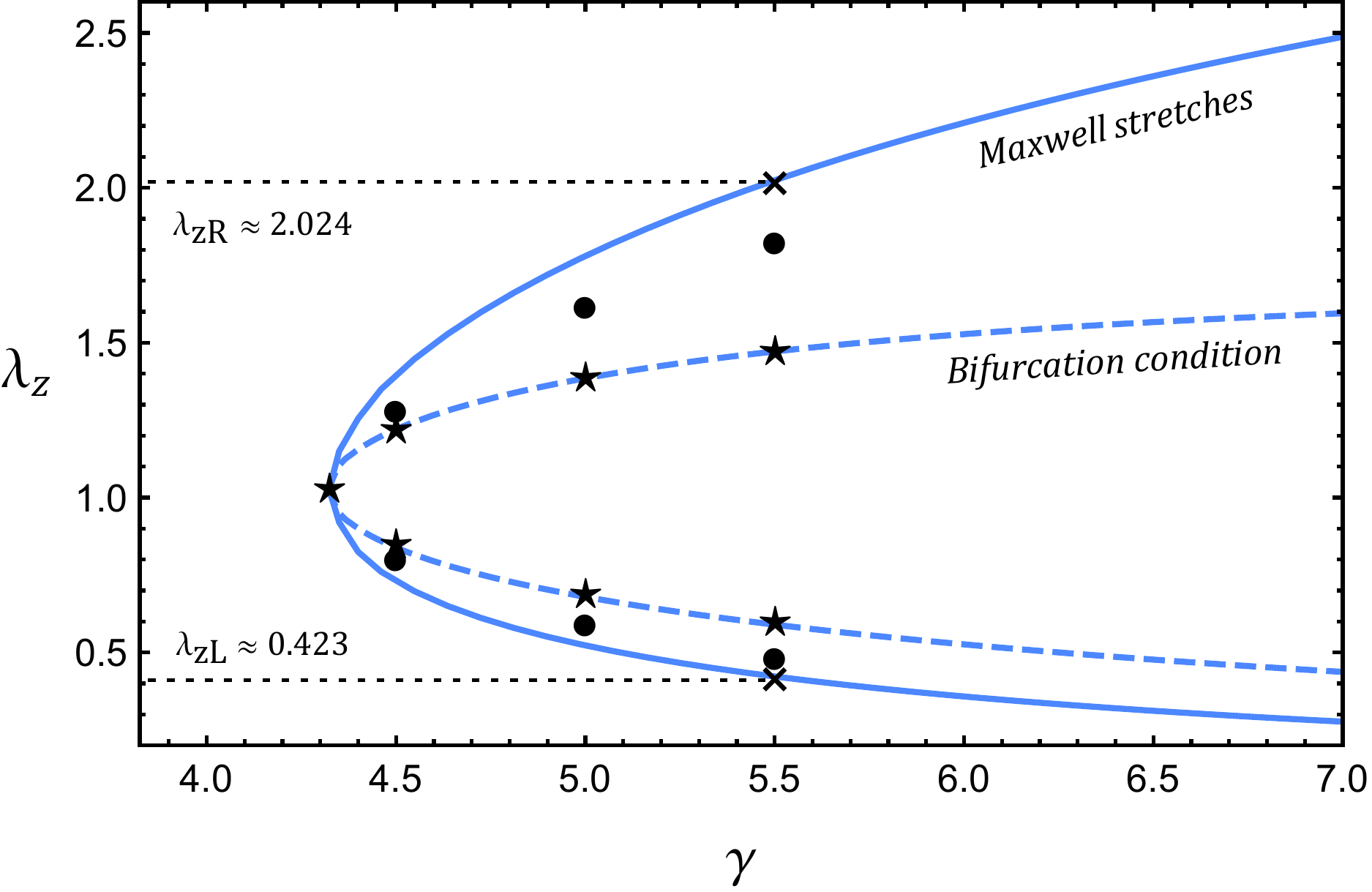}};
\node at (7.5,0) {\includegraphics[scale=0.37]{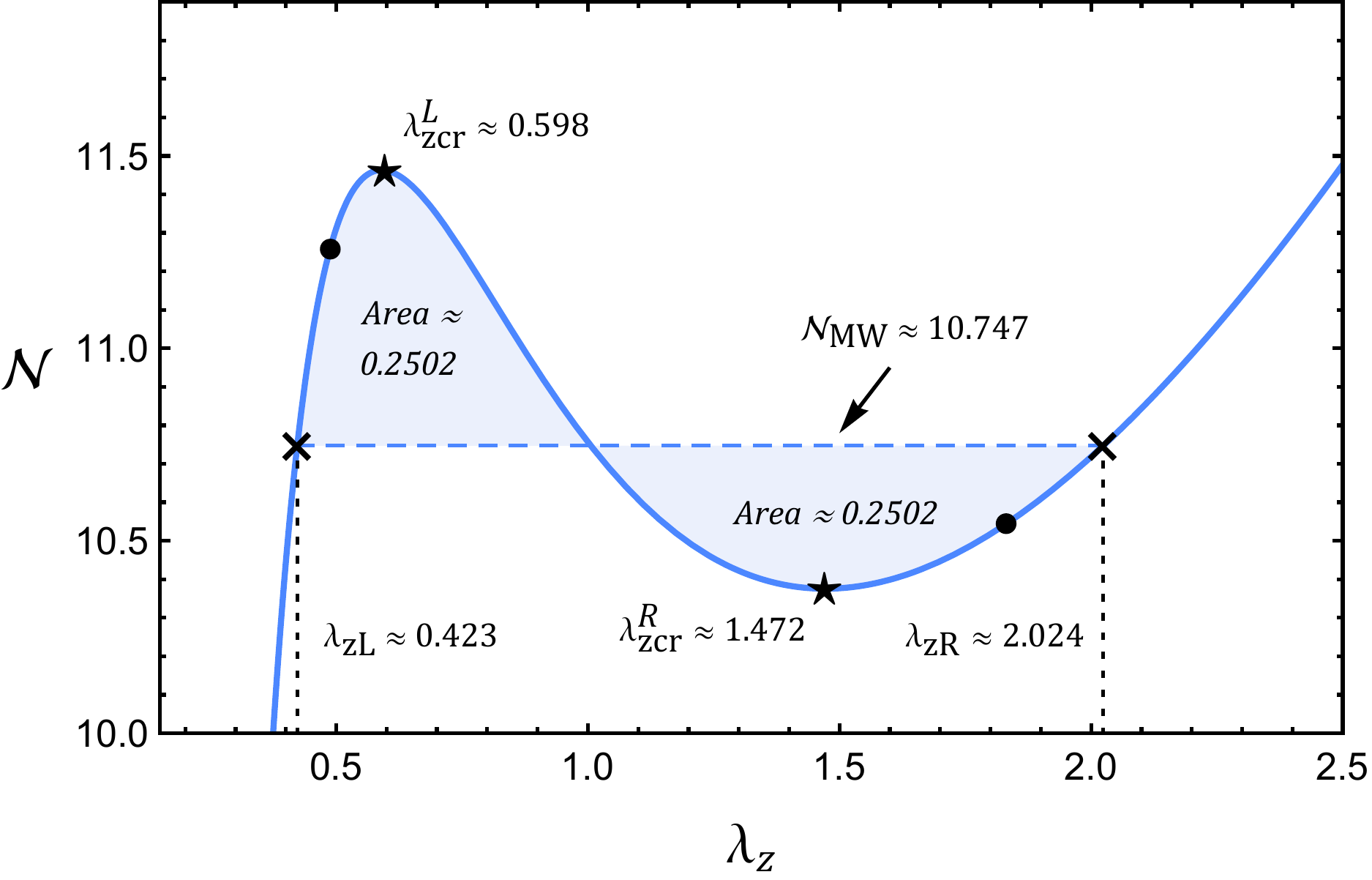}};
\node at (0,-5.5) {\includegraphics[scale=0.365]{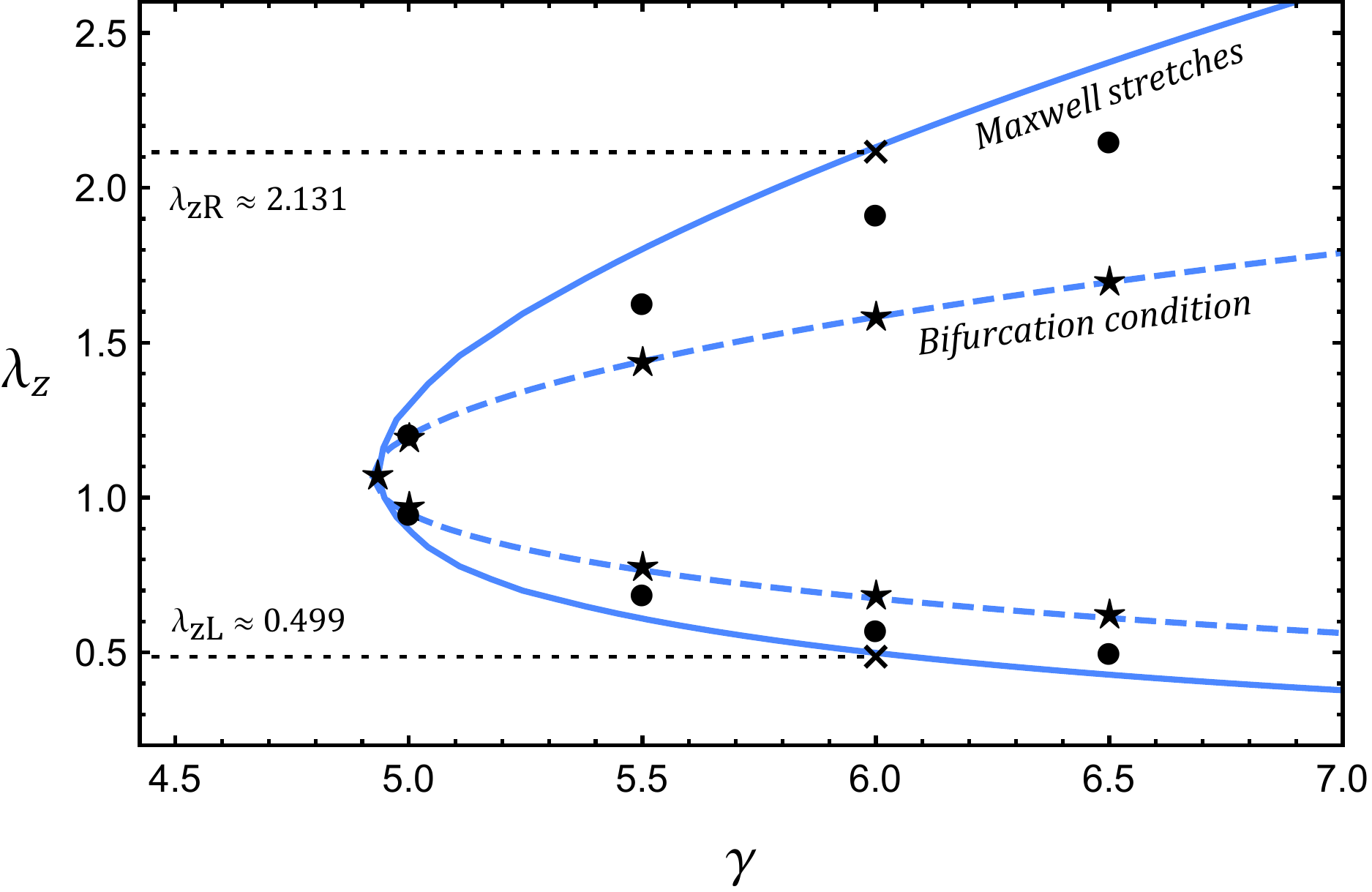}};
\node at (7.5,-5.5) {\includegraphics[scale=0.37]{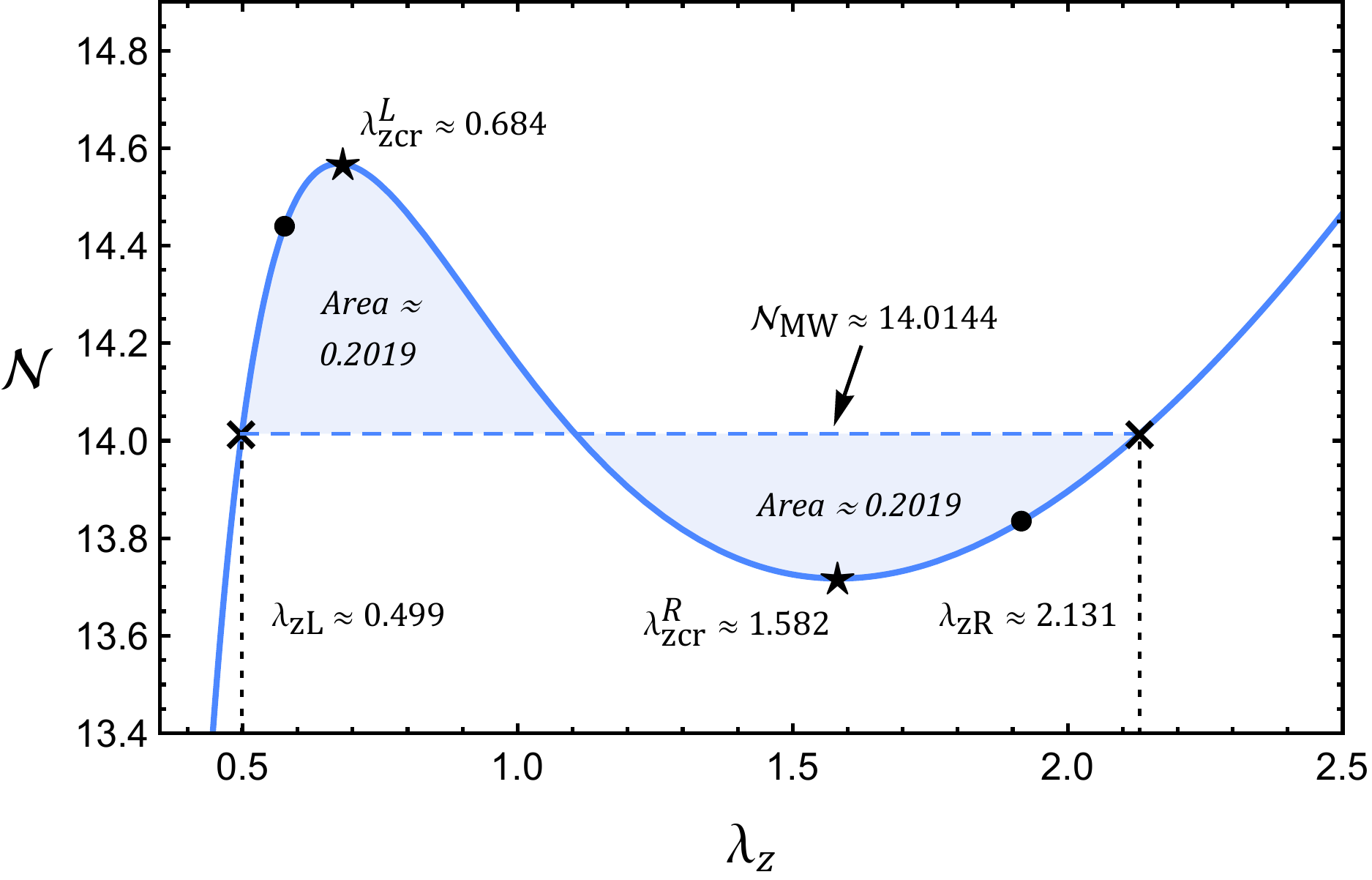}};
\node[text width=1cm] at (0.5,2.75)
    {\footnotesize \text{(a)}};
\node[text width=1cm] at (8.1,2.75)
    {\footnotesize \text{(b)}};
\node[text width=1cm] at (0.5,-2.75)
    {\footnotesize \text{(c)}};
\node[text width=1cm] at (8.1,-2.75)
    {\footnotesize \text{(d)}};
\end{tikzpicture}
\caption{In (a) and (c), the critical stretches $\lzcr$ from our theoretical bifurcation condition (dashed blue curve) and the Maxwell stretches $\lambda_{zL}$ and $\lambda_{zR}$ computed using the equal area rule (solid blue curve) are presented in the $(\lz,\gamma)$ plane for $\nu=0.4$, and the quadratic and logarithmic neo-Hookean material models, respectively. The black stars and dots give the numerical simulation results from DJ for the bifurcation points and the Maxwell stretches, respectively. In (b) and (d), we superpose our theoretical bifurcation points and Maxwell stretches as well as the numerically determined values of DJ on the $\mc{N}=\mc{N}(\lz)$ curve for $\gamma=6$. This demonstrates that the theoretically determined Maxwell stretches satisfy the equal area rule, whereas the numerically simulated stretches don't.}
\label{PBB}
\end{figure}
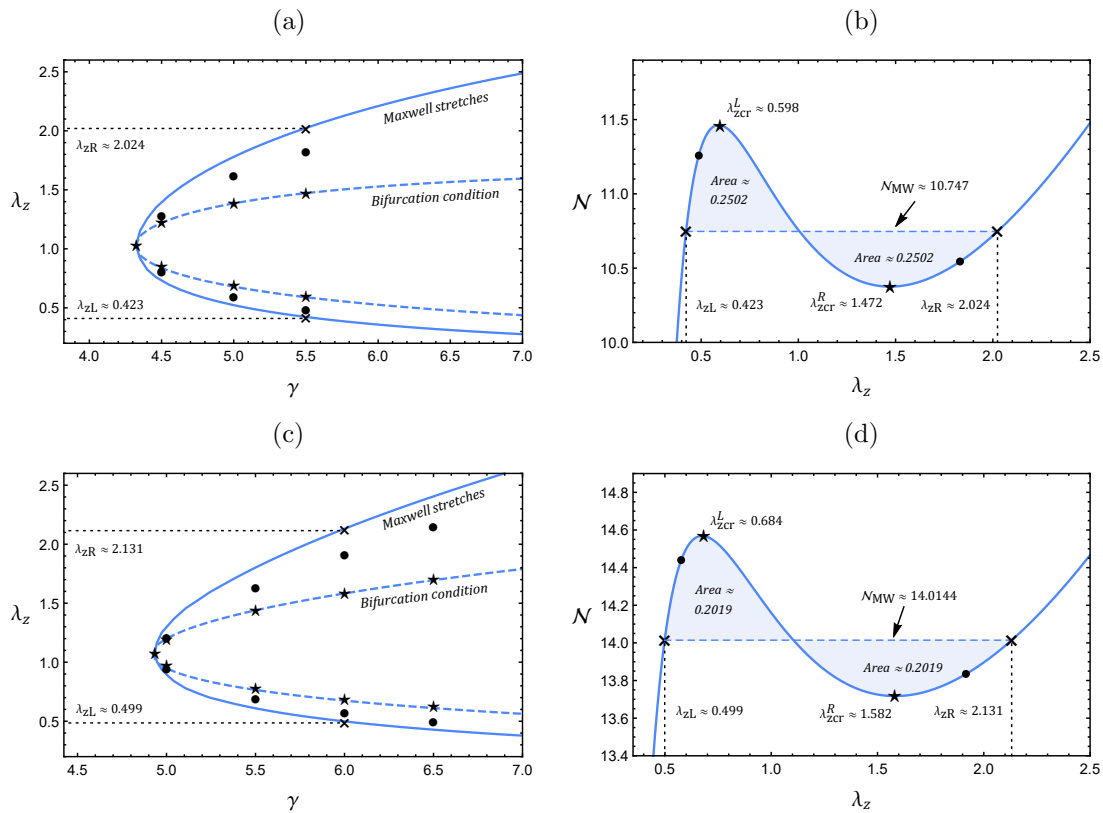

\begin{figure}[h!]
\centering
\includegraphics[scale=0.4]{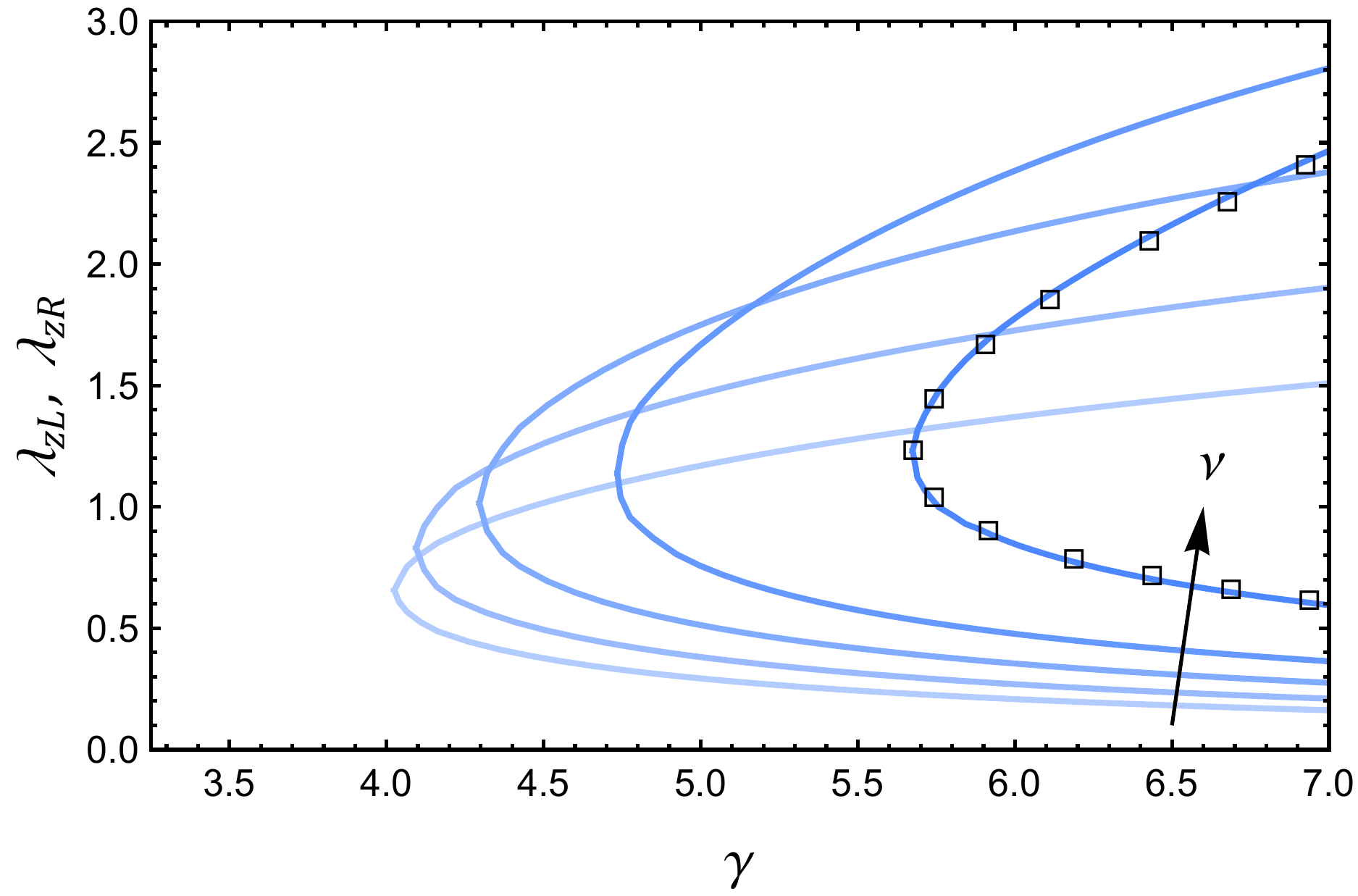}
\caption{Plots of the Maxwell stretches $\lambda_{zL}$ and $\lambda_{zR}$ (determined from the equal area rule) against $\gamma$ for the compressible Gent material model with $\Jm=100$ and $\nu=0.05,0.3,0.4,0.45,0.499$. The black squares give the associated results from \cite{FuST} for the incompressible case, and the arrow indicates the direction of parameter growth.}
\label{MWGENT}
\end{figure}

We now turn our attention to the compressible cylinder case. In Fig. $\ref{PBB}$ (a) and (c), we present theoretical results for the bifurcation values of $\lz$ (dashed curve) and the Maxwell stretches (solid curve) as functions of $\gamma$ for the quadratic and logarithmic neo-Hookean models, respectively. The black stars are numerically simulated bifurcation points taken from Fig. $11$ (d) of DJ, and we observe perfect agreement with our theory. However, the black dots, which are the numerically simulated Maxwell stretches for different values of $\gamma$ from DJ, are at odds with our theoretical predictions. The basic principle of Maxwell's equal area rule in the present context is that the stretches $\lambda_{zL}$ and $\lambda_{zR}$ should be defined such that the magnitude of the areas between the horizontal line passing through the points $(\lambda_{zL},\mc{N}_{MW})$ and $(\lambda_{zR},\mc{N}_{MW})$, and the curve $\mc{N}=\mc{N}(\lz)$ above and below the horizontal line, should be equal. Only if this condition is satisfied can a ``two-phase" state exist. We show in Fig. $\ref{PBB}$ (b) and (d) that this requirement is satisfied by our Maxwell stretches, but not by those predicted in DJ. Thus, it is clear that the analytical approach presented is a powerful tool in guiding numerical simulation studies of elastic phase-separation-like phenomena, and in validating the results of such studies. In Fig. $\ref{MWGENT}$, we plot the Maxwell stretches $\lambda_{zL}$ and $\lambda_{zR}$ as functions of $\gamma$ obtained from the equal area rule for the compressible Gent material model $(\ref{gentC})$. We set $\Jm=100$ and fix $\nu$ at several different values. The right-most curve corresponds to $\nu\RA 1/2$ (i.e. the incompressible limit), and to provide validation of our theoretical results, we compare with the corresponding FEM simulation results presented in Fig. 13 (a) of \cite{FuST} (black squares). We observe that there is exceptional agreement between the two sets of results, emphasising the point that the equal area rule should act as a consistency check for numerical simulation results (and vice versa).

We highlight the remarkable capability of the equal area rule in describing the post-bifurcation behaviour of the compressible cylinder using only analytical expressions for force parameters corresponding to the primary deformation. Not only does the approach yield the Maxwell stretches for each portion of the fully developed ``two-phase" state, it can also fully predict the nature of the axial propagation of the bulged or depressed phase. For example, in the fixed $\lz$ and increasing $\gamma$ scenario covered previously, say we fix $\lz>\lambda_{\tx{min}}$. Then, we understand that a localized bulge will emerge at $\gcr$, followed by a snap-through to a ``two-phase" state comprising of an axially propagated bulged ``phase" with stretch $\lambda_{zL}$ and a depressed ``phase" with stretch $\lambda_{zR}$. The longitudinal proportion of the bulged ``phase" with respect to the overall ``two-phase" state can easily be computed through $\lambda_{zL}/\lz$. Now, from our results in Fig. $\ref{MWGENT}$, we observe that $\lambda_{zL}$ is generally a decreasing function of $\gamma$. Thus, as we increase $\gamma$ beyond its bifurcation value, the \textit{proportion} of the propagated bulge ``phase" with respect to the overall length of the cylinder will also decrease. In other words, an increasing surface tension in the fully non-linear regime will cause a reduction in the axial length of the bulged ``phase". We note, however, that $d\lambda_{zL}/d\gamma$ is typically small, and tends to zero as $\gamma$ gets sufficiently large. Hence, the reduction of ``bulged" phase length as $\gamma>\gcr$ is increased is very gradual, and when $\gamma$ gets large enough, the length of the bulged ``phase" will seemingly approach a non-zero limiting value.

\section{Concluding remarks}
The complete bifurcation behaviour of an incompressible solid cylinder under axial loading and surface tension is fully understood. However, with the exception of the numerical study of DJ, analogous studies when the cylinder is \textit{compressible} are scarce. In this paper, we have provided greater theoretical insights into the bifurcation behaviour of compressible solid cylinders and a source of comparison for existing and future numerical studies. By drawing upon known results for the mathematically similar problem of localized bulging in hollow tubes under axial loading and internal inflation, we derived analytical bifurcation conditions for elasto-capillary localized bulging or necking in compressible solid cylinders for three distinct loading scenarios. For the quadratic and logarithmic neo-Hookean material models, we found perfect agreement between our theoretical bifurcation conditions and the numerical simulation conditions presented in DJ. 

Based on the findings for an incompressible solid cylinder and hollow tube, we then conjectured that, when the axial stretch $\lz$ is fixed and the surface tension $\gamma$ is increased beyond its bifurcation value, the tube undergoes a phase-separation-like process where the initial localized solution evolves into a ``two-phase" state. We explained how, with the use of the simple analytical expression $(\ref{compN})$ for $\mc{N}$, the Maxwell stretches $\lambda_{zL}$ and $\lambda_{zR}$ associated with this ``two-phase" state can be determined as functions of $\gamma$ through the equal area rule. On comparing the results of the equal area rule approach with the corresponding numerical simulation results in \cite{FuST} and DJ, we found perfect agreement in the former case but disagreement in the latter. This highlighted that numerical studies of phase-separation-like phenomena don't always use the equal area rule as a consistency check on their results, and it is hoped that this paper will invoke change in this regard in future studies.

\subsection*{Acknowledgements}
The author thanks Prof. Yibin Fu from Keele University for several helpful discussions, and acknowledges the School of Computing and Mathematics, Keele University for funding their PhD studies through a faculty scholarship.

\bibliographystyle{elsarticle-harv}
\biboptions{authoryear}

\begin{thebibliography}{36}
\expandafter\ifx\csname natexlab\endcsname\relax\def\natexlab#1{#1}\fi
\providecommand{\url}[1]{\texttt{#1}}
\providecommand{\href}[2]{#2}
\providecommand{\path}[1]{#1}
\providecommand{\DOIprefix}{doi:}
\providecommand{\ArXivprefix}{arXiv:}
\providecommand{\URLprefix}{URL: }
\providecommand{\Pubmedprefix}{pmid:}
\providecommand{\doi}[1]{\href{http://dx.doi.org/#1}{\path{#1}}}
\providecommand{\Pubmed}[1]{\href{pmid:#1}{\path{#1}}}
\providecommand{\bibinfo}[2]{#2}
\ifx\xfnm\relax \def\xfnm[#1]{\unskip,\space#1}\fi
\bibitem[{{Abaqus}(2013)}]{ab2013}
\bibinfo{author}{{Abaqus}}, \bibinfo{year}{2013}.
\newblock \bibinfo{title}{ABAQUS Analysis Users Manual, version 6.13}.
\newblock \bibinfo{note}{Dassault Systems, Providence, RI, USA}.
\bibitem[{Althobaiti(2022)}]{althobaiti2022}
\bibinfo{author}{Althobaiti, A.}, \bibinfo{year}{2022}.
\newblock \bibinfo{title}{Effect of torsion on the initiation of localized
  bulging in a hyperelastic tube of arbitrary thickness}.
\newblock \bibinfo{journal}{Z. Angew. Math. Phys.} \bibinfo{volume}{73},
  \bibinfo{pages}{1--11}.
\bibitem[{Bar-Ziv and Moses(1994)}]{bar1994}
\bibinfo{author}{Bar-Ziv, R.}, \bibinfo{author}{Moses, E.},
  \bibinfo{year}{1994}.
\newblock \bibinfo{title}{Instability and" pearling" states produced in tubular
  membranes by competition of curvature and tension}.
\newblock \bibinfo{journal}{Phys. Rev. Lett.} \bibinfo{volume}{73},
  \bibinfo{pages}{1392}.
\bibitem[{Bevilacqua et~al.(2020)Bevilacqua, Shao, Saylor, Bostwick and
  Ciarletta}]{bevilacqua2020}
\bibinfo{author}{Bevilacqua, G.}, \bibinfo{author}{Shao, X.},
  \bibinfo{author}{Saylor, J.R.}, \bibinfo{author}{Bostwick, J.B.},
  \bibinfo{author}{Ciarletta, P.}, \bibinfo{year}{2020}.
\newblock \bibinfo{title}{Faraday waves in soft elastic solids}.
\newblock \bibinfo{journal}{Proc. R. Soc. A} \bibinfo{volume}{476},
  \bibinfo{pages}{20200129}.
\bibitem[{Bico et~al.(2018)Bico, Reyssat and Roman}]{bico2018}
\bibinfo{author}{Bico, J.}, \bibinfo{author}{Reyssat, {\'E}.},
  \bibinfo{author}{Roman, B.}, \bibinfo{year}{2018}.
\newblock \bibinfo{title}{Elastocapillarity: When surface tension deforms
  elastic solids}.
\newblock \bibinfo{journal}{Annu. Rev. Fluid Mech.} \bibinfo{volume}{50},
  \bibinfo{pages}{629--659}.
\bibitem[{Carew et~al.(1968)Carew, Vaishnav and Patel}]{carew1968}
\bibinfo{author}{Carew, T.E.}, \bibinfo{author}{Vaishnav, R.N.},
  \bibinfo{author}{Patel, D.J.}, \bibinfo{year}{1968}.
\newblock \bibinfo{title}{Compressibility of the arterial wall}.
\newblock \bibinfo{journal}{Cir. Res.} \bibinfo{volume}{23},
  \bibinfo{pages}{61--68}.
\bibitem[{Clerk-Maxwell(1875)}]{JCM1875}
\bibinfo{author}{Clerk-Maxwell, J.}, \bibinfo{year}{1875}.
\newblock \bibinfo{title}{On the dynamical evidence of the molecular
  constitution of bodies}.
\newblock \bibinfo{journal}{J. Chem. Soc} \bibinfo{volume}{28},
  \bibinfo{pages}{493--508}.
\bibitem[{Datar et~al.(2019)Datar, Ameeramja, Bhat, Srivastava, Mishra, Bernal,
  Prost, Callan-Jones and Pullarkat}]{datar2019}
\bibinfo{author}{Datar, A.}, \bibinfo{author}{Ameeramja, J.},
  \bibinfo{author}{Bhat, A.}, \bibinfo{author}{Srivastava, R.},
  \bibinfo{author}{Mishra, A.}, \bibinfo{author}{Bernal, R.},
  \bibinfo{author}{Prost, J.}, \bibinfo{author}{Callan-Jones, A.},
  \bibinfo{author}{Pullarkat, P.A.}, \bibinfo{year}{2019}.
\newblock \bibinfo{title}{The roles of microtubules and membrane tension in
  axonal beading, retraction, and atrophy}.
\newblock \bibinfo{journal}{Biophys. J.} \bibinfo{volume}{117},
  \bibinfo{pages}{880--891}.
\bibitem[{Dortdivanlioglu and Javili(2022)}]{dort2022}
\bibinfo{author}{Dortdivanlioglu, B.}, \bibinfo{author}{Javili, A.},
  \bibinfo{year}{2022}.
\newblock \bibinfo{title}{Plateau rayleigh instability of soft elastic solids.
  effect of compressibility on pre and post bifurcation behavior}.
\newblock \bibinfo{journal}{Extreme Mech. Lett.} \bibinfo{volume}{55},
  \bibinfo{pages}{101797}.
\bibitem[{Emery and Fu(2021a)}]{emery2021MOSM}
\bibinfo{author}{Emery, D.R.}, \bibinfo{author}{Fu, Y.B.},
  \bibinfo{year}{2021}a.
\newblock \bibinfo{title}{Elasto-capillary circumferential buckling of soft
  tubes under axial loading: existence and competition with localised beading
  and periodic axial modes}.
\newblock \bibinfo{journal}{Mech. Soft Mater.} \bibinfo{volume}{3}.
\newblock \bibinfo{note}{Doi: https://doi.org/10.1007/s42558-021-00034-x}.
\bibitem[{Emery and Fu(2021b)}]{emery2021IJSS}
\bibinfo{author}{Emery, D.R.}, \bibinfo{author}{Fu, Y.B.},
  \bibinfo{year}{2021}b.
\newblock \bibinfo{title}{Localised bifurcation in soft cylindrical tubes under
  axial stretching and surface tension}.
\newblock \bibinfo{journal}{Int. J. Solids Struct.} \bibinfo{volume}{219},
  \bibinfo{pages}{23--33}.
\bibitem[{Emery and Fu(2021c)}]{emery2021PRSA}
\bibinfo{author}{Emery, D.R.}, \bibinfo{author}{Fu, Y.B.},
  \bibinfo{year}{2021}c.
\newblock \bibinfo{title}{Post-bifurcation behaviour of elasto-capillary
  necking and bulging in soft tubes}.
\newblock \bibinfo{journal}{arXiv preprint arXiv:2104.04713} .
\bibitem[{Fong et~al.(1999)Fong, Chun and Reneker}]{fong1999}
\bibinfo{author}{Fong, H.}, \bibinfo{author}{Chun, I.},
  \bibinfo{author}{Reneker, D.H.}, \bibinfo{year}{1999}.
\newblock \bibinfo{title}{Beaded nanofibers formed during electrospinning}.
\newblock \bibinfo{journal}{Polymer} \bibinfo{volume}{40},
  \bibinfo{pages}{4585--4592}.
\bibitem[{Fu et~al.(2018)Fu, Dorfmann and Xie}]{fu2018}
\bibinfo{author}{Fu, Y.B.}, \bibinfo{author}{Dorfmann, L.},
  \bibinfo{author}{Xie, Y.}, \bibinfo{year}{2018}.
\newblock \bibinfo{title}{Localized necking of a dielectric membrane}.
\newblock \bibinfo{journal}{Extreme Mech. Lett.} \bibinfo{volume}{21},
  \bibinfo{pages}{44--48}.
\bibitem[{Fu et~al.(2021)Fu, Jin and Goriely}]{FuST}
\bibinfo{author}{Fu, Y.B.}, \bibinfo{author}{Jin, L.},
  \bibinfo{author}{Goriely, A.}, \bibinfo{year}{2021}.
\newblock \bibinfo{title}{Necking, beading, and bulging in soft elastic
  cylinders.}
\newblock \bibinfo{journal}{J. Mech. Phys. Solids} \bibinfo{volume}{147},
  \bibinfo{pages}{104250}.
\bibitem[{Fu et~al.(2016)Fu, Liu and Francisco}]{fu2016}
\bibinfo{author}{Fu, Y.B.}, \bibinfo{author}{Liu, J.L.},
  \bibinfo{author}{Francisco, G.S.}, \bibinfo{year}{2016}.
\newblock \bibinfo{title}{Localized bulging in an inflated cylindrical tube of
  arbitrary thickness--the effect of bending stiffness}.
\newblock \bibinfo{journal}{J. Mech. Phys. Solids} \bibinfo{volume}{90},
  \bibinfo{pages}{45--60}.
\bibitem[{Fu et~al.(2008)Fu, Pearce and Liu}]{fu2008}
\bibinfo{author}{Fu, Y.B.}, \bibinfo{author}{Pearce, S.P.},
  \bibinfo{author}{Liu, K.K.}, \bibinfo{year}{2008}.
\newblock \bibinfo{title}{Post-bifurcation analysis of a thin-walled
  hyperelastic tube under inflation}.
\newblock \bibinfo{journal}{Int. J. Non-Linear Mech.} \bibinfo{volume}{43},
  \bibinfo{pages}{697--706}.
\bibitem[{Giudici and Biggins(2020)}]{giudici2020}
\bibinfo{author}{Giudici, A.}, \bibinfo{author}{Biggins, J.S.},
  \bibinfo{year}{2020}.
\newblock \bibinfo{title}{Ballooning, bulging and necking: an exact solution
  for longitudinal phase separation in elastic systems near a critical point}.
\newblock \bibinfo{journal}{Phys. Rev. E} \bibinfo{volume}{102},
  \bibinfo{pages}{033007}.
\bibitem[{Haughton and Ogden(1979a)}]{ho1979a}
\bibinfo{author}{Haughton, D.M.}, \bibinfo{author}{Ogden, R.W.},
  \bibinfo{year}{1979}a.
\newblock \bibinfo{title}{Bifurcation of inflated circular cylinders of elastic
  material under axial loading—i. membrane theory for thin-walled tubes}.
\newblock \bibinfo{journal}{J. Mech. Phys. Solids} \bibinfo{volume}{27},
  \bibinfo{pages}{179--212}.
\bibitem[{Haughton and Ogden(1979b)}]{ho1979b}
\bibinfo{author}{Haughton, D.M.}, \bibinfo{author}{Ogden, R.W.},
  \bibinfo{year}{1979}b.
\newblock \bibinfo{title}{Bifurcation of inflated circular cylinders of elastic
  material under axial loading—ii. exact theory for thick-walled tubes}.
\newblock \bibinfo{journal}{J. Mech. Phys. Solids} \bibinfo{volume}{27},
  \bibinfo{pages}{489--512}.
\bibitem[{Kilinc et~al.(2009)Kilinc, Gallo and Barbee}]{kilinc2009}
\bibinfo{author}{Kilinc, D.}, \bibinfo{author}{Gallo, G.},
  \bibinfo{author}{Barbee, K.A.}, \bibinfo{year}{2009}.
\newblock \bibinfo{title}{Interactive image analysis programs for quantifying
  injury-induced axonal beading and microtubule disruption}.
\newblock \bibinfo{journal}{Comput. Methods Programs Biomed.}
  \bibinfo{volume}{95}, \bibinfo{pages}{62--71}.
\bibitem[{Kyriakides and Yu-Chung(1990)}]{kyriakides1990}
\bibinfo{author}{Kyriakides, S.}, \bibinfo{author}{Yu-Chung, C.},
  \bibinfo{year}{1990}.
\newblock \bibinfo{title}{On the inflation of a long elastic tube in the
  presence of axial load}.
\newblock \bibinfo{journal}{Int. J. Solids Struct.} \bibinfo{volume}{26},
  \bibinfo{pages}{975--991}.
\bibitem[{Liu et~al.(2019)Liu, Ye, Althobaiti and Xie}]{liu2019}
\bibinfo{author}{Liu, Y.}, \bibinfo{author}{Ye, Y.},
  \bibinfo{author}{Althobaiti, A.}, \bibinfo{author}{Xie, Y.X.},
  \bibinfo{year}{2019}.
\newblock \bibinfo{title}{Prevention of localized bulging in an inflated
  bilayer tube}.
\newblock \bibinfo{journal}{Int. J. Mech. Sci.} \bibinfo{volume}{153},
  \bibinfo{pages}{359--368}.
\bibitem[{Mallock(1891)}]{mallock1891}
\bibinfo{author}{Mallock, A.}, \bibinfo{year}{1891}.
\newblock \bibinfo{title}{Ii. note on the instability of india-rubber tubes and
  balloons when distended by fluid pressure}.
\newblock \bibinfo{journal}{Proc. R. Soc.} \bibinfo{volume}{49},
  \bibinfo{pages}{458--463}.
\bibitem[{Matsuo and Tanaka(1992)}]{matsuo1992}
\bibinfo{author}{Matsuo, E.S.}, \bibinfo{author}{Tanaka, T.},
  \bibinfo{year}{1992}.
\newblock \bibinfo{title}{Patterns in shrinking gels}.
\newblock \bibinfo{journal}{Nature} \bibinfo{volume}{358},
  \bibinfo{pages}{482--485}.
\bibitem[{Pandey et~al.(2021)Pandey, Kansal, Herrada, Eggers and
  Snoeijer}]{pandey2021}
\bibinfo{author}{Pandey, A.}, \bibinfo{author}{Kansal, M.},
  \bibinfo{author}{Herrada, M.A.}, \bibinfo{author}{Eggers, J.},
  \bibinfo{author}{Snoeijer, J.H.}, \bibinfo{year}{2021}.
\newblock \bibinfo{title}{Elastic rayleigh--plateau instability: dynamical
  selection of nonlinear states}.
\newblock \bibinfo{journal}{Soft matter} \bibinfo{volume}{17},
  \bibinfo{pages}{5148--5161}.
\bibitem[{Riccobelli(2021)}]{riccobelli2021}
\bibinfo{author}{Riccobelli, D.}, \bibinfo{year}{2021}.
\newblock \bibinfo{title}{Active elasticity drives the formation of periodic
  beading in damaged axons}.
\newblock \bibinfo{journal}{Phys. Rev. E} \bibinfo{volume}{104},
  \bibinfo{pages}{024417}.
\bibitem[{Taffetani and Ciarletta(2015a)}]{taffetani2015}
\bibinfo{author}{Taffetani, M.}, \bibinfo{author}{Ciarletta, P.},
  \bibinfo{year}{2015}a.
\newblock \bibinfo{title}{Beading instability in soft cylindrical gels with
  capillary energy: weakly non-linear analysis and numerical simulations}.
\newblock \bibinfo{journal}{J. Mech. Phys. Solids} \bibinfo{volume}{81},
  \bibinfo{pages}{91--120}.
\bibitem[{Taffetani and Ciarletta(2015b)}]{taffetani2015a}
\bibinfo{author}{Taffetani, M.}, \bibinfo{author}{Ciarletta, P.},
  \bibinfo{year}{2015}b.
\newblock \bibinfo{title}{Elastocapillarity can control the formation and the
  morphology of beads-on-string structures in solid fibers}.
\newblock \bibinfo{journal}{Phys. Rev. E} \bibinfo{volume}{91},
  \bibinfo{pages}{032413}.
\bibitem[{Wang et~al.(2017)Wang, Althobaiti and Fu}]{wang2017}
\bibinfo{author}{Wang, J.}, \bibinfo{author}{Althobaiti, A.},
  \bibinfo{author}{Fu, Y.B.}, \bibinfo{year}{2017}.
\newblock \bibinfo{title}{Localized bulging of rotating elastic cylinders and
  tubes}.
\newblock \bibinfo{journal}{J. Mech. Mater. Struct.} \bibinfo{volume}{12},
  \bibinfo{pages}{545--561}.
\bibitem[{Wang and Fu(2018)}]{wang2018effect}
\bibinfo{author}{Wang, J.}, \bibinfo{author}{Fu, Y.B.}, \bibinfo{year}{2018}.
\newblock \bibinfo{title}{Effect of double-fibre reinforcement on localized
  bulging of an inflated cylindrical tube of arbitrary thickness}.
\newblock \bibinfo{journal}{J. Eng. Math.} \bibinfo{volume}{109},
  \bibinfo{pages}{21--30}.
\bibitem[{Wang et~al.(2021)Wang, Liu, Wang, Chen and Wu}]{wang2021large}
\bibinfo{author}{Wang, Q.}, \bibinfo{author}{Liu, M.}, \bibinfo{author}{Wang,
  Z.}, \bibinfo{author}{Chen, C.}, \bibinfo{author}{Wu, J.},
  \bibinfo{year}{2021}.
\newblock \bibinfo{title}{Large deformation and instability of soft hollow
  cylinder with surface effects}.
\newblock \bibinfo{journal}{J. Appl. Mech.} \bibinfo{volume}{88}.
\bibitem[{{Wolfram Research Inc.}(2021)}]{wo2021}
\bibinfo{author}{{Wolfram Research Inc.}}, \bibinfo{year}{2021}.
\newblock \bibinfo{title}{Mathematica 12.3.1}.
\newblock \URLprefix \url{https://www.wolfram.com/mathematica}.
  \bibinfo{note}{champaign, IL}.
\bibitem[{Xuan and Biggins(2016)}]{xuan2016}
\bibinfo{author}{Xuan, C.}, \bibinfo{author}{Biggins, J.},
  \bibinfo{year}{2016}.
\newblock \bibinfo{title}{Finite-wavelength surface-tension-driven
  instabilities in soft solids, including instability in a cylindrical channel
  through an elastic solid}.
\newblock \bibinfo{journal}{Phys. Rev. Lett.} \bibinfo{volume}{94},
  \bibinfo{pages}{023107}.
\bibitem[{Xuan and Biggins(2017)}]{xuan2017}
\bibinfo{author}{Xuan, C.}, \bibinfo{author}{Biggins, J.},
  \bibinfo{year}{2017}.
\newblock \bibinfo{title}{Plateau-rayleigh instability in solids is a simple
  phase separation}.
\newblock \bibinfo{journal}{Phys. Rev. E} \bibinfo{volume}{95},
  \bibinfo{pages}{053106}.
\bibitem[{Yu and Fu(2022)}]{yu2022}
\bibinfo{author}{Yu, X.}, \bibinfo{author}{Fu, Y.B.}, \bibinfo{year}{2022}.
\newblock \bibinfo{title}{An analytic derivation of the bifurcation conditions
  for localization in hyperelastic tubes and sheets}.
\newblock \bibinfo{journal}{Z. Angew Math. Phys.} \bibinfo{volume}{73},
  \bibinfo{pages}{1--16}.

\end{thebibliography}

\end{document}